\documentclass[11pt]{article}
\usepackage{amsmath}
\usepackage{amssymb}
\usepackage{amsthm}
\usepackage{amsfonts}
\usepackage{comment}
\usepackage{color}
\usepackage{mathrsfs}
\usepackage{braket}
\usepackage{graphicx}
\usepackage[subrefformat=parens]{subcaption}
\usepackage{cite}
\usepackage{here}

\usepackage[stable]{footmisc}

\makeatletter

\@addtoreset{equation}{section}
\makeatother

\begin{document}

\pagestyle{empty}

\begin{flushright}
\end{flushright}

\vspace{3cm}

\begin{center}

{\bf\LARGE  
Violation of bound on chaos for charged probe in Kerr-Newman-AdS black hole 
}
\\

\vspace*{1.5cm}
{\large 
Bogeun Gwak$^\clubsuit$\footnote{rasenis@dgu.ac.kr}, Naoto Kan$^\clubsuit$\footnote{naotokan000@gmail.com}, Bum-Hoon Lee$^{\spadesuit\diamondsuit}$\footnote{bhl@sogang.ac.kr} and Hocheol Lee$^\diamondsuit$\footnote{insaying@sogang.ac.kr}
} \\
\vspace*{0.5cm}

{\it 
$^\clubsuit$Division of Physics and Semiconductor Science, Dongguk University, Seoul 04620, Republic of Korea
\\
$^\spadesuit$Center for Quantum Spacetime, Sogang University, Seoul 04107, Republic of Korea
\\
$^\diamondsuit$Department of Physics, Sogang University, Seoul 04107, Republic of Korea
}

\end{center}

\vspace*{1.0cm}

\begin{abstract}
{\noindent
We investigate the conjectured bound on the Lyapunov exponent for a charged particle with angular motion in the Kerr-Newman-AdS black hole. The Lyapunov exponent is calculated based on the effective Lagrangian. We show that the negative cosmological constant reduces the chaotic behavior of the particle, namely, it decreases the Lyapunov exponent. Hence, the bound is more effective in the AdS spacetime than in the flat spacetime. Nevertheless, we find that the bound can be violated when the angular momenta of the black hole are turned on. Moreover, we show that in an extremal black hole, the bound is more easily violated compared to the nonextremal black hole.

}
\end{abstract}

\newpage
\baselineskip=18pt
\setcounter{page}{2}
\pagestyle{plain}
\baselineskip=18pt
\pagestyle{plain}

\setcounter{footnote}{0}
	
\section{Introduction}
The anti-de Sitter/conformal field theory (AdS/CFT) correspondence \cite{Maldacena:1997re} has been well studied over the last several decades.
The first proposal of this correspondence was deduced from string theory.
The gravity side of the correspondence is described by a theory on the AdS space, which is a solution of a supergravity theory in the near-horizon region.
The field theory side is described by the supersymmetric gauge theory, which is the low-energy effective theory on D-branes.
In the AdS/CFT correspondence, the strong coupling region of the CFT side corresponds to the weak coupling limit of the AdS side.
Therefore, nonperturbative quantities in CFT can be calculated using perturbative quantities in AdS via the AdS/CFT correspondence.
For example, we can calculate correlation functions of primary operators in the CFT by using the GKP-Witten relation \cite{Gubser:1998bc,Witten:1998qj}.
We can also apply the AdS/CFT correspondence to finite temperature systems \cite{Witten:1998qj,Witten:1998zw}.
The gravity side is then described by black hole geometries with a negative cosmological constant, i.e., AdS black holes.
AdS black holes can exhibit rotation as well as asymptotically-flat black holes.
Indeed, AdS black holes with rotation have been studied in the context of the AdS/CFT correspondence \cite{Hawking:1998kw}.

Chaos theory has been extensively studied since the start of the twentieth century.
A system is referred to as chaotic when the difference between the initial values of the system grows exponentially.
The sensitivity to the initial values is quantitively evaluated using the Lyapunov exponent.
Suppose that $X^i(t)$ is a vector in phase space, the Lyapunov exponent is defined as $|\partial X^i(t)/\partial X^j(0)|\sim e^{\lambda t}$.
The difference between the initial values increases exponentially when the Lyapunov exponent $\lambda$ is positive.
This discussion is based on chaos theory in classical dynamics.
Chaos theory in quantum systems has also been investigated.
Based on canonical quantization, it was proposed that the chaos in thermal quantum systems could be characterized based on the large degree of freedom, $N\gg 1 $, by $C(t)=-\braket{[V(t),W(0)}_\beta$ \cite{Larkin:1969qu}, where $V(t)$ and $W(t)$ are any Hermitian operators in the quantum systems, and $\braket{\cdots}_\beta$ denotes the average over the thermal ensemble.
$\beta$ is the inverse temperature.
There are two time scales for the behavior of $C(t)$: the dissipation time $t_d\sim \beta$, and the scrambling time or the Ehrenfest time $t_\ast\sim \beta \log N$.
Before the dissipation time, $C(t)$ is almost zero, and after the scrambling time, $C(t)$ is saturated.
An interesting behavior of $C(t)$ is observed after the dissipation time and before the scrambling time.
When the system is chaotic, $C(t)$ exponentially increases between $t_d$ and $t_\ast$, $C(t)\sim N^{-2}e^{2\lambda_L t}$, where $\lambda_L$ is referred to as the quantum Lyapunov exponent.

The chaos in the quantum systems has been studied using the AdS/CFT correspondence.
In \cite{Shenker:2013pqa}, the chaotic behavior was discussed based on the dual gravity system.
They considered a system with a Banados-Teitelboim-Zanelli (BTZ) black hole \cite{Banados:1992wn}, which is a three-dimensional asymptotically-AdS black hole.
The field theory introduced a small perturbation on the boundary as described by the BTZ black hole with a shock wave \cite{Aichelburg:1970dh,Dray:1984ha,Hotta:1992qy} in the gravity side.
Sensitivity to the initial condition in the case of the boundary theory was observed via the AdS/CFT correspondence.
Maldacena, Shenker, and Stanford conjectured the existence of the universal bound on the quantum Lyapunov exponent \cite{Maldacena:2015waa},
		\begin{align}
		\lambda_L\le \frac{2\pi T}{\hbar},
	\label{eq:bound}
	\end{align}
where we suppose $k_B=1$.
This bound was studied in many systems.
For example, in the Sachdev-Ye-Kitaev (SYK) model, it was confirmed that the bound is saturated.

Chaotic behavior is closely related to the nonlinearity of dynamics.
It is well known that general relativity includes nonlinearity.
In particular, the Lyapunov exponent of a probe around a black hole has been studied extensively \cite{Dettmann:1994dj,Suzuki:1996gm,Suzuki:1999si,Han:2008zzf,Verhaaren:2009md,PandoZayas:2010xpn,Pradhan:2012rkk,Pradhan:2012qf,Basu:2012ae,Pradhan:2013bli,Pradhan:2014tva,Jawad:2016kgt,Lukes-Gerakopoulos:2016xoc,Chen:2016tmr,Hashimoto:2016wme,Jai-akson:2017ldo,Dalui:2018qqv,Li:2018wtz,Nunez:2018ags,Hashimoto:2018fkb,Akutagawa:2018yoe,Akutagawa:2019awh,Cubrovic:2019qee,Giataganas:2021ghs,Lei:2021koj} (and also \cite{Barrow:1981sx,Bombelli:1991eg,Yurtsever:1994yb,Letelier:1997uv,deMoura:1999wf,Setare:2010zd,Lukes-Gerakopoulos:2016udm,Liu:2017fjx,Zelenka:2019nyp,Yi:2020shw,Mondal:2021exj}).
Recently, Hashimoto and Tanahashi applied the bound on the Lyapunov exponent \eqref{eq:bound} to a black hole system using a probe particle \cite{Hashimoto:2016dfz}.
The temperature of the system was given by the Hawking temperature. As such, the right-hand side of \eqref{eq:bound} is equal to the surface gravity $\kappa$.
In \cite{Hashimoto:2016dfz}, the authors considered the effective Lagrangian with a static gauge, which is described by an inverse harmonic oscillator.
The Lyapunov exponent was calculated using the effective Lagrangian.
They showed that when the interaction of the particle is given by the electromagnetic force or the scalar force, the bound on the Lyapunov exponent is saturated.
The bound on the Lyapunov exponent was also investigated in this context for other static and spherically symmetric black holes \cite{Zhao:2018wkl,Lei:2020clg}.
The violation of the bound was found for several cases.

In this work, we investigate the bound on the Lyapunov exponent for a charged particle with angular motion in a Kerr-Newman-AdS (KN-AdS) black hole.
The rotating black hole was considered by \cite{Kan:2021blg,Yu:2022ysm}, but they did not use the cosmological constant.
Particularly, we consider the negative cosmological constant, so the boundary of the spacetime is AdS. Since the bound on the Lyapunov exponent is based on the AdS/CFT correspondence, chaotic behavior becomes more meaningful in AdS black holes with angular momentum. 
We will show that the negative cosmological constant plays a significant role in terms of validity and violation on the bound in a rotating black hole.
The contribution of the angular momentum and electric charge of a particle to the Lyapunov exponent will be discussed when the negative cosmological constant is present.
Finally, the cosmological constant makes the particle less chaotic, but the bound on the Lyapunov exponent can be still violated for specific parameters. 

The remainder of this report is organized into several sections. 
In Section 2, we review the KN-AdS black hole and provide the KN-AdS metric and several conditions for parameters.
In Section 3, we calculate the effective Lagrangian for a charged probe particle around a KN-AdS black hole.
We also calculate the (maximum) Lyapunov exponent using the effective Lagrangian.
In Section 4, we analyze an RN-AdS black hole.
We also evaluate the role of the cosmological constant in the presence of the angular motion of the particle and
algebraically consider several limits of the RN-AdS black hole.  
In Section 5, we consider the bound of a rotating black hole with a negative cosmological constant, i.e., Kerr-AdS and KN-AdS black holes.
Finally, Section 5 summarizes the main conclusions.

\section{Review of the Kerr-Newman-AdS black hole}
\noindent
In this section, we review the basics of KN-AdS black holes.
At the end of this section, we will arrive at an expression for the bound \eqref{eq:bound}.
The metric of the KN-AdS black hole is given by
	\begin{align}
	\label{eq:Kerr-Newman AdS black hole}
		ds^2&=-\frac{\Delta_r}{\rho^2}\left(dt-\frac{a\sin^2\theta}{\Xi}d\phi\right)^2+\frac{\rho^2}{\Delta_r}dr^2+\frac{\rho^2}{\Delta_\theta}d\theta^2+\frac{\Delta_\theta\sin^2\theta}{\rho^2}\left(adt-\frac{r^2+a^2}{\Xi}d\phi\right)^2 \\
		&=-\frac{1}{\rho^2}\left(\Delta_r-a^2\Delta_\theta\sin^2\theta\right)dt^2+\frac{2a\sin^2\theta}{\Xi \rho^2}\left(\Delta_r-(r^2+a^2)\Delta_\theta\right)dtd\phi \nonumber \\
		&\qquad +\frac{\sin^2\theta}{\Xi^2\rho^2}\left((r^2+a^2)^2\Delta_\theta-a^2\Delta_r\sin^2\theta\right)d\phi^2+\frac{\rho^2}{\Delta_r	}dr^2+\frac{\rho^2}{\Delta_\theta}d\theta^2,
	\end{align}
where 
	\begin{gather}
		\rho^2=r^2+a^2\cos^2\theta, \\
		\Xi=1+\frac{a^2}{3}\Lambda \\
		\Delta_r=(r^2+a^2)\left(1-\frac{r^2}{3}\Lambda\right)-2Mr+Q^2, \\
		\Delta_\theta=1+\frac{a^2}{3}\Lambda\cos^2\theta.
	\end{gather}
We assumed that the magnetic charge is zero.
The electromagnetic potential is 
	\begin{align}
		A=-\frac{Qr}{\rho^2}\left(dt-\frac{a\sin^2\theta}{\Xi}d\phi\right).
	\label{eq:EM_potential}
	\end{align}
$Q$ is the electric charge, $a$ is the rotating parameter of the black hole, and $\Lambda$ is the (negative) cosmological constant.
Note that the metric is valid only for $a^2<-3\Lambda^{-1}$; it becomes singular when $a^2=-3\Lambda^{-1}$.
In the limit of $a^2=-3\Lambda^{-1}$, the three-dimensional Einstein universe at infinity rotates at the speed of light. 
To see this, we define the angular velocity of the rotating Einstein universe at infinity:
	\begin{align}
		\Omega=\Omega_{\rm H}-\Omega_\infty,
	\end{align}
where $\Omega_{\rm H}$ and $\Omega_\infty$ are the angular velocity at the horizon and infinity, respectively.
These are given by
	\begin{align}
		\Omega_{\rm H}&=\frac{a\Xi}{r_+^2+a^2}, \\
		\Omega_\infty &=\frac{a}{3}\Lambda.
	\end{align}
Then, the angular velocity of the rotating Einstein universe becomes
	\begin{align}
		\Omega=-\frac{a}{3}\frac{\Lambda r_+^2-3	}{r_+^2+a^2}.
	\end{align}
In the limit $a^2=-3\Lambda^{-1}$, we have $\Omega=\sqrt{-\Lambda/3}$.

In the coordinates $(t,r,\phi,\theta)$, the metric \eqref{eq:Kerr-Newman AdS black hole} does not seem to be the asymptotically AdS space at infinity.
This is because the observer at infinity is not static.
To resolve this problem, we introduce a coordinate transformation \cite{Hawking:1998kw},
	\begin{align}
		t\to T,\qquad \phi\to \Phi+\frac{1}{3}a\Lambda T.
	\end{align}
Under this coordinate transformation, the metric becomes
	\begin{align}
		ds^2&=-\frac{\Delta_r}{\rho^2\Xi^2}\left(\Delta_\theta dT-a\sin^2\theta d\Phi\right)^2+\frac{\rho^2}{\Delta_r}dr^2+\frac{\rho^2}{\Delta_\theta}d\theta^2+\frac{\Delta_\theta \sin^2\theta}{\rho^2\Xi^2}\left\{a\left(1-\frac{r^2}{3}\Lambda\right)dT-(r^2+a^2)d\Phi\right\}^2 \\
		&=-\frac{\Delta_\theta}{\rho^2\Xi^2}\left\{\Delta_r\Delta_\theta-a^2\left(1-\frac{r^2}{3}\Lambda\right)^2\sin^2\theta\right\}dT^2+\frac{2a\Delta_\theta\sin^2\theta}{\rho^2\Xi^2}\left\{\Delta_r-(r^2+a^2)\left(1-\frac{r^2}{3}\Lambda\right)\right\}dTd\Phi \nonumber \\
		&\qquad +\frac{\sin^2\theta}{\rho^2\Xi^2}\left\{\Delta_\theta(r^2+a^2)^2-a^2\Delta_r\sin^2\theta\right\}d\Phi^2+\frac{\rho^2}{\Delta_r}dr^2+\frac{\rho^2}{\Delta_\theta}d\theta^2,
	\label{eq:KN_metric_static}
	\end{align}
and the vector potential is given by
	\begin{align}
		A=-\frac{Qr}{\rho^2\Xi}\left(\Delta_\theta dT-a\sin^2\theta d\Phi\right).
	\end{align}

It is convenient to define the critical mass parameter $M_{\rm extr}$ \cite{Caldarelli:1998hg}:
	\begin{align}
		M_{\rm extr}=&\frac{1}{3}\sqrt{-\frac{1}{2\Lambda}}\left(\sqrt{\left(1-\frac{a^2}{3}\Lambda\right)^2-4\Lambda(a^2+Q^2)}-\frac{2a^2}{3}\Lambda+2\right) \nonumber \\
		&\qquad\qquad\quad\times\left(\sqrt{\left(1-\frac{a^2}{3}\Lambda\right)^2-4\Lambda(a^2+Q^2)}+\frac{a^2}{3}\Lambda-1\right)^{1/2}.
	\end{align}
For $M<M_{\rm extr}$, we have naked singularities.
For $M>M_{\rm extr}$, the outer event horizon is larger than the inner Cauchy horizon.
For $M=M_{\rm extr}$, the lapse function $\Delta_r$ has a double root, and the metic \eqref{eq:Kerr-Newman AdS black hole} describes the extremal black hole.
Thus we assume $M\ge M_{\rm extr}$ and $a^2<-3\Lambda^{-1}$ in the following.

The event horizon $r_+$ of the black hole is the largest solution of $\Delta_r=0$.
The Hawking temperature $T_{\rm H}$ is given by
	\begin{align}
		T_{\rm H}=\frac{r_+\left(1-\frac{a^2}{3}\Lambda-r_+^2\Lambda-\frac{a^2+Q^2}{r_+^2}\right)}{4\pi \left(r_+^2+a^2\right)}.
	\label{eq:Hawking_temperature}
	\end{align}
The surface gravity is given by $\kappa=2\pi T_H/\hbar$, so the bound on the Lyapunov exponent \eqref{eq:bound} can be expressed as
	\begin{align}
		\lambda\le \kappa.
	\end{align}
We will also use the squared expression; $\lambda^2\le \kappa^2$

\section{The effective Lagrangian and the Lyapunov exponent in the Kerr-Newman-AdS black hole}
\noindent
Let us consider a probe particle around a KN-AdS black hole.
We will see that the motion of the particle is effectively described by an (inverse) harmonic oscillator near a local extremum of the effective potential.
Then, we can determine the Lyapunov exponent for the effective Lagrangian.

We begin with the Polyakov-type action of a particle on curved space.
The action of the probe charged particle is given by
	\begin{gather}
		S=\int d\tau {\cal L} \\
		{\cal L}=\frac{1}{2e}g_{\mu\nu}\dot{X}^\mu\dot{X}^\nu-\frac{e}{2}m^2+qA_\mu\dot{X}^\mu,
	\end{gather}
where $m$ and $q$ are the mass and charge of the particle, respectively.
The field $e$ is the auxiliary field.
This action is equivalent to the Nambu-Goto-type action by integrating out the auxiliary field.

We choose the static gauge, $\tau=T$.
The Lagrangian with the metric \eqref{eq:KN_metric_static} is given by
	\begin{align}
		{\cal L}&=\frac{1}{2e}\Bigg[-\frac{\Delta_\theta}{\rho^2\Xi^2}\left\{\Delta_r\Delta_\theta-a^2\left(1-\frac{r^2}{3}\Lambda\right)^2\sin^2\theta\right\}+\frac{2a\Delta_\theta\sin^2\theta}{\rho^2\Xi^2}\left\{\Delta_r-(r^2+a^2)\left(1-\frac{r^2}{3}\Lambda\right)\right\}\dot{\Phi} \nonumber \\
		&\qquad+\frac{\sin^2\theta}{\rho^2\Xi^2}\left\{\Delta_\theta(r^2+a^2)^2-a^2\Delta_r\sin^2\theta\right\}\dot{\Phi}^2+\frac{\rho^2}{\Delta_r}\dot{r}^2+\frac{\rho^2}{\Delta_\theta}\dot{\theta}^2\Bigg] \nonumber \\
		&\qquad-\frac{e}{2}m^2-q\frac{Qr\Delta_\theta}{\rho^2\Xi}+q\frac{Qar\sin^2\theta}{\rho^2\Xi}\dot{\Phi},
	\end{align}
where the dots represent the derivative of $\tau$.
Let us focus on the equatorial motion of the particle, $\theta=\pi/2$.
As such, we have $\rho^2=r^2$ and $\Delta_\theta=1$.
Thus, the Lagrangian for the equatorial motion is 
	\begin{align}
		{\cal L}&=\frac{1}{2e}\Bigg[-\frac{1}{r^2\Xi^2}\left\{\Delta_r-a^2\left(1-\frac{r^2}{3}\Lambda\right)^2\right\}+\frac{2a}{r^2\Xi^2}\left\{\Delta_r-(r^2+a^2)\left(1-\frac{r^2}{3}\Lambda\right)\right\}\dot{\Phi} \nonumber \\
		&\qquad+\frac{1}{r^2\Xi^2}\left\{(r^2+a^2)^2-a^2\Delta_r\right\}\dot{\Phi}^2+\frac{r^2}{\Delta_r}\dot{r}^2\Bigg]-\frac{e}{2}m^2-q\frac{Q}{r\Xi}+q\frac{Qa}{r\Xi}\dot{\Phi}.
	\end{align}
	
The Lagrangian is invariant under the constant shift of the $\Phi$-direction, $\Phi\to\Phi+\alpha$.
We can define the conserved angular momentum of the particle:
	\begin{align}
		L=\frac{\partial{\cal L}}{\partial\dot{\Phi}}=\frac{a \left(a^2 \left(\Lambda  r^2-3\right)+3 \Delta +3 e \Xi  q Q r+\Lambda  r^4-3 r^2\right)+3 \dot{\Phi}\left(\left(a^2+r^2\right)^2-a^2 \Delta \right)}{3 e \Xi ^2 r^2}.
	\end{align}
In addition, based on the equation of motion for $e$, $\dot{X}^\mu\dot{X}_\mu=-e^2m^2$, 
we have
	\begin{align}
		e^2=\frac{r^2}{\alpha(r)}\left(\Delta_r-\frac{\beta(r)}{\Delta_r}\dot{r}^2\right)
	\end{align}
where
	\begin{gather}
		\alpha(r)=a^4 m^2+a^2 \left(m^2 \left(2 r^2-\Delta \right)+q^2 Q^2\right)-2 a L \Xi  q Q r+L^2 \Xi ^2 r^2+m^2 r^4, \\
		\beta(r)=\left(a^2+r^2\right)^2-a^2 \Delta_r.
	\end{gather}

Let us consider the effective motion of the particle.
The effective Lagrangian is typically defined as
	\begin{align}
		{\cal L}_{\rm eff}=&{\cal L}-L\dot{\Phi}
	\end{align}
using the conserved angular momentum.
We will focus on the motion of the particle near a local extremum of the effective potential.
To analyze the sensitivity to the initial condition of the particle, we suppose that the initial velocity of the particle is small.
Then, the particle near the local extremum of the potential is described by the nonrelativistic limit, $\dot{r}^2\ll1$, and the effective Lagrangian is given by
	\begin{align}
		{\cal L}_{\rm eff}=\frac{1}{2}K(r)\dot{r}^2-V_{\rm eff}(r)+O(\dot{r}^4),
	\label{eq:eff_Lagrangian}
	\end{align}
where
	\begin{align}
		K(r)=\frac{r\sqrt{\alpha(r)}}{\Delta_r^{3/2}},
	\end{align}
and 
	\begin{align}
		V_{\rm eff}(r)=&\frac{1}{3\Xi \beta(r)}\bigg(-a L \Xi  \left(a^2 \left(\Lambda  r^2-3\right)+3 \Delta_r +\Lambda  r^4-3 r^2\right) \nonumber \\
		&\qquad\quad+q Q r \left(a^2 \Lambda +3\right) \left(a^2+r^2\right)+r\Delta_r^{1/2}\left(a^2\Lambda+3\right)\sqrt{\alpha(r)}\bigg).
	\label{eq:eff_potential}
	\end{align}

To evaluate the maximum of the Lyapunov exponent, we focus on the motion of the particle near the local extremum.
We denote the position of the local extremum as $r_0$, which is obtained by solving $V_{\rm eff}'(r_0)=0$.
For a small perturbation $r(t)=r_0+\epsilon(t)$, the effective Lagrangian is approximately described by
	\begin{align}
		{\cal L}_{\rm eff}=\frac{1}{2}K(r_0)\left(\dot{\epsilon^2}+\lambda^2\epsilon^2\right),
	\end{align}
where we eliminate a constant term.
$\lambda^2$ in the second term is given by
	\begin{align}
		\lambda^2=\left.-\frac{V''_{\rm eff}(r)}{K(r)}\right|_{r=r_0},
	\end{align}
where 
	\begin{align}
		V_{\rm eff}''(r)=&-\frac{\beta '(r)}{3 \Xi  \Delta_r^{1/2} \alpha (r)^{1/2} \beta (r)^2}\bigg[2 \sqrt{\Delta_r\,\alpha (r)} \Big(-2 a L \Xi  r \left(a^2 \Lambda +2 \Lambda  r^2-3\right) \nonumber \\
		&\qquad\qquad\qquad\qquad\qquad\quad+q Q \left(a^2   \Lambda +3\right) \left(a^2+3 r^2\right)-3 a L \Xi  \Delta_r'\Big) \nonumber \\
		&\qquad\qquad\qquad\qquad\qquad\quad+r \left(a^2 \Lambda +3\right) \Delta_r\alpha '(r)+\left(a^2 \Lambda +3\right) \alpha (r) \left(r \Delta_r'+2 \Delta_r\right)\bigg] \nonumber \\
		&+\frac{2 \beta '(r)^2-\beta (r) \beta ''(r)}{3\Xi  \beta (r)^3}\bigg[\left(a^2+r^2\right) \left(q Q r \left(a^2 \Lambda +3\right)+a L \Xi  \left(3-\Lambda r^2\right)\right) \nonumber \\
		&\qquad\qquad\qquad\qquad\qquad\qquad+r \left(a^2 \Lambda +3\right) \sqrt{\Delta_r\,\alpha (r)} -3 a L \Xi  \Delta_r\bigg] \nonumber \\
		&+\frac{1}{12 \Xi \Delta_r^{3/2} \alpha (r)^{3/2} \beta (r) }\bigg[-4 a L \Xi  \Delta_r^{3/2} \alpha (r)^{3/2} \left(2 a^2 \Lambda +12 \Lambda  r^2+3 \Delta_r''-6\right) \nonumber \\
		&\qquad\qquad\qquad\qquad\qquad\quad+24 q Q r \left(a^2 \Lambda +3\right)\Delta_r^{3/2}  \alpha (r)^{3/2} \nonumber \\
		&\qquad\qquad\qquad\qquad\qquad\quad+4 \left(a^2 \Lambda +3\right) \alpha (r) \Delta_r\left(\Delta_r \alpha'(r)+\Delta_r'\alpha (r) \right) \nonumber \\
		&\qquad\qquad\qquad\qquad\qquad\quad+r \left(a^2 \Lambda +3\right) \big(2 \alpha (r) \Delta_r \left(\alpha '(r) \Delta_r'+\Delta_r''\alpha (r) \right) \nonumber \\
		&\qquad\qquad\qquad\qquad\qquad\quad-\Delta_r^2\left(\alpha '(r)^2-2 \alpha (r) \alpha ''(r)\right)-\Delta_r'^2\alpha (r)^2 \big)\bigg].
	\end{align}
When $r_0$ is the position of the local maximum, we have $V''_{\rm eff}|_{r=r_0}<0$, i.e., $\lambda^2>0$.
This is the case when the effective motion of the particle is described by an inverse harmonic oscillator, which implies that the divergence of the close trajectories grows exponentially.
This behavior is associated with chaos (e.g., see \cite{Hashimoto:2016wme,Morita:2021syq,Morita:2021mfi,Bhattacharyya:2020art}), and (the square of) the Lyapunov exponent is given by $\lambda^2$.
We also note that the effective motion is described by a harmonic oscillator when $V''_{\rm eff}|_{r=r_0}>0$, i.e., $\lambda^2<0$, thus, we do not have chaos.

\section{The bound for the RN-AdS black hole}
In the previous section, we obtained the Lyapunov exponent for a particle near the local maximum.
Using this result, we investigate the bound on the Lyapunov exponent,
	\begin{align}
		\kappa^2-\lambda^2\ge 0
	\end{align}
for the RN-AdS black hole in this section.

\subsection{Numerical analyses}
We first numerically analyze the bound of the Lyapunov exponent for a massless and massive particle.
In the numerical analysis, we use dimensionless parameters; we set $M=1$.
The results for a massless particle are shown in Fig. \ref{fig:m=0_RN}, which includes six plots.
For each plot, the horizontal axis represents the particle’s charge $q$, and the vertical axis represents the particle’s angular momentum $L$.
In the case of an RN-AdS black hole, the angular momentum of the particle appears as the square in the effective potential. Therefore, we consider only the positive region of the angular momentum $L$.
The plots in the same row are the results of the same black hole charge but for different cosmological constants.
The leftmost plots in each row show the asymptotically-flat black holes, and the rightmost plots show the AdS black holes with $\Lambda=1$, which is a large cosmological constant.
The plots in the same columns are the results for the same cosmological constant but different black hole charges.
The top plots in each column show the Schwarzschild-type (noncharged) black holes, and the bottom plots show the extremal (maximally charged) black holes.
$Q_{\rm Max}$ represents the extremal charge of each black hole, which is given by $Q_{\rm Max}=1.00, 0.945, 0.912$ for black holes with $\Lambda=0, -0.5, -1$, respectively.
For example, the upper left plot is for the asymptotically-flat Schwarzschild black hole, and the lower right plot is for the extremal RN-AdS black hole with $\Lambda=1$.
In Fig. \ref{fig:m=1_RN}, we also show the results for a massive particle with $m=1$.
The plots are arranged in the same way as in Fig. \ref{fig:m=0_RN}.

In Fig. \ref{fig:m=0_RN} and \ref{fig:m=1_RN}, the colored regions (the sets of the colored points) denote that the square of the Lyapunov exponent is positive $\lambda^2>0$, namely, the particle exhibits chaotic behavior, and the bound of the Lyapunov exponent is violated $\kappa^2-\lambda^2<0$.
The color denotes the value of $\kappa^2-\lambda^2$.
The gray regions (the sets of gray points) indicate that the square of the Lyapunov exponent is positive $\lambda^2>0$, but the bound is satisfied $\kappa^2-\lambda^2\ge 0$.
The white regions do not exhibit chaos; $\lambda^2<0$ and $\kappa^2-\lambda^2\ge 0$.

The middle and bottom plots in the left column in Fig. \ref{fig:m=1_RN} are the results for the nonextremal and extremal asymptotically-flat RN black holes for a massive particle, respectively.
These cases have been studied by \cite{Kan:2021blg}, and the results obtained are consistent with their findings.
For all plots, including the AdS black hole case, the regions with $L=0$ (q-axis) are white. Thus, the bound is satisfied, which was also established by \cite{Zhao:2018wkl}.
Turning on $L$, almost all the regions are colored or gray.
This implies that the angular momentum of the particle makes the particle more chaotic.

To observe the effect of the cosmological constant, let us focus on the middle row in Fig. \ref{fig:m=1_RN}.
In the plot of the asymptotically-flat black hole (the leftmost plot), the bound is violated in the almost parameter region.
However, if we turn on the cosmological constant, it is evident that size of the violation region (the colored region) decreases, and also the size of the chaotic region (combined the colored and gray region) decreases.
This implies that the cosmological constant suppresses the chaotic behavior of particles.
However, the bound is still violated in the extremal black hole.
The bound for the extremal black hole can be more easily violated compared to the nonextremal case.
We can also see the same suppression in the other rows and in Fig. \ref{fig:m=0_RN}.

\begingroup
\renewcommand{\arraystretch}{5}
\begin{figure}[htbp]
	\begin{tabular}{ccc}
		\begin{minipage}[t]{0.32\hsize}
		\centering
			\includegraphics[width=46mm]{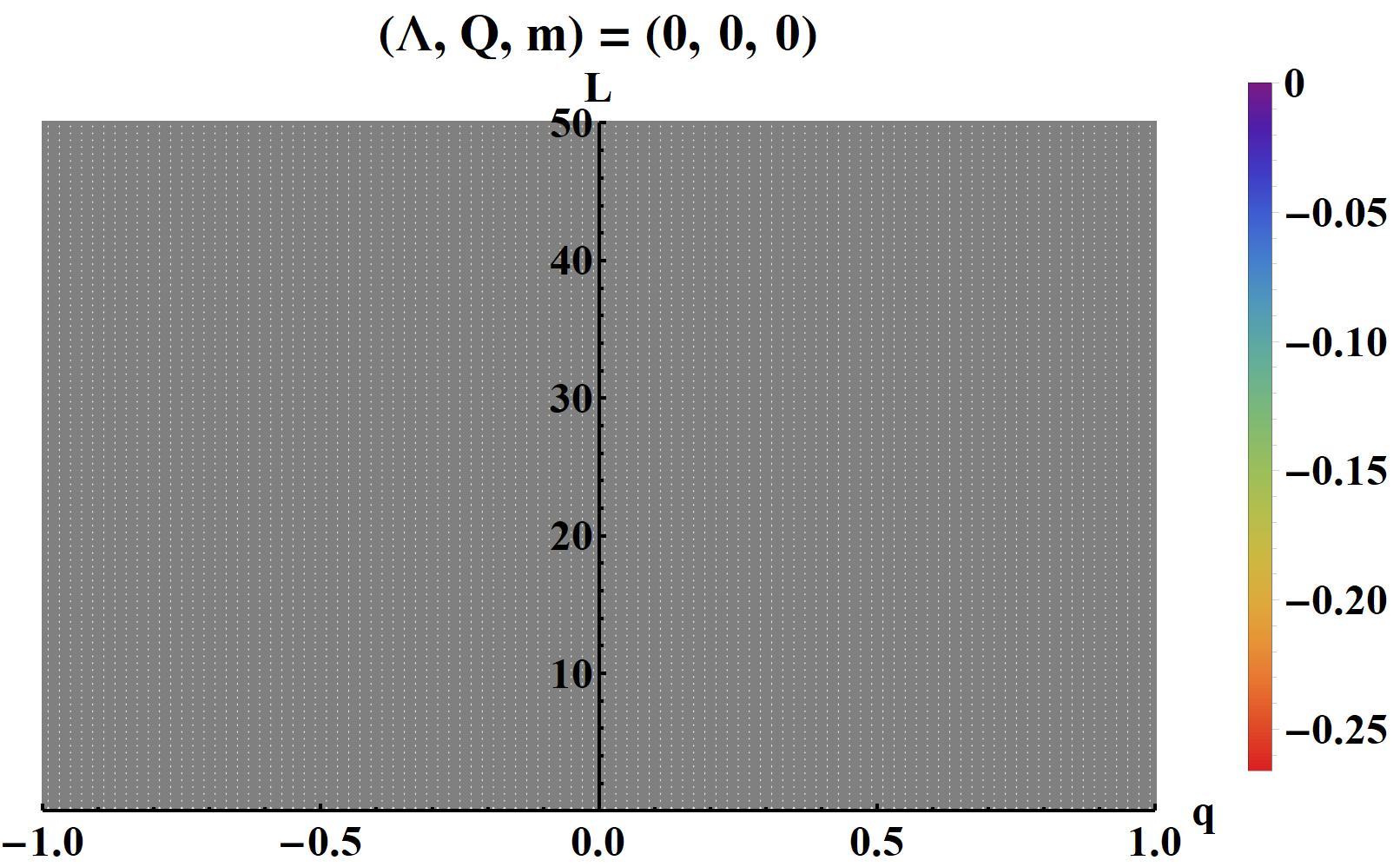}
		\end{minipage} &
		\begin{minipage}[t]{0.32\hsize}
		\centering
			\includegraphics[width=46mm]{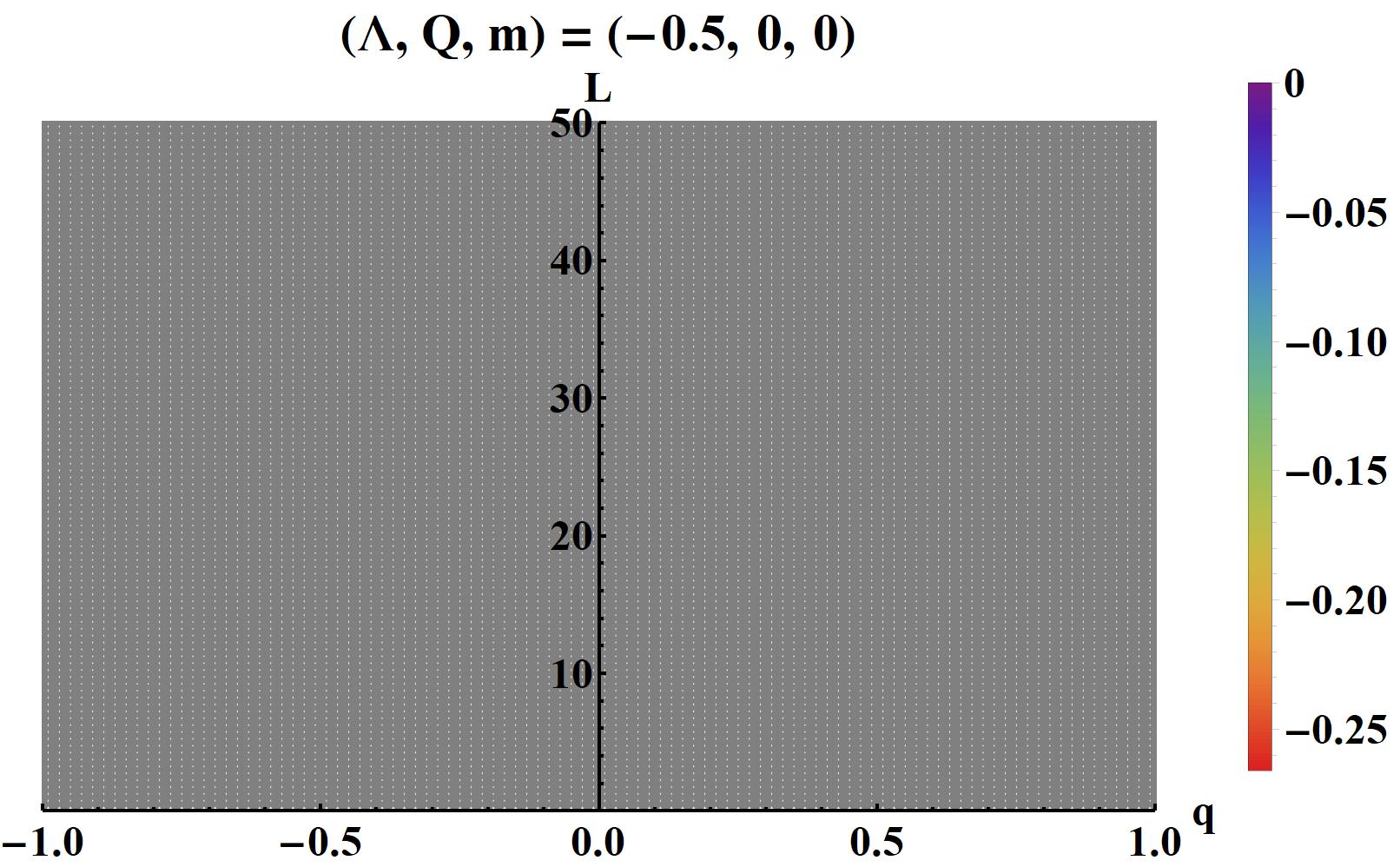}
		\end{minipage} &
		\begin{minipage}[t]{0.32\hsize}
		\centering
			\includegraphics[width=46mm]{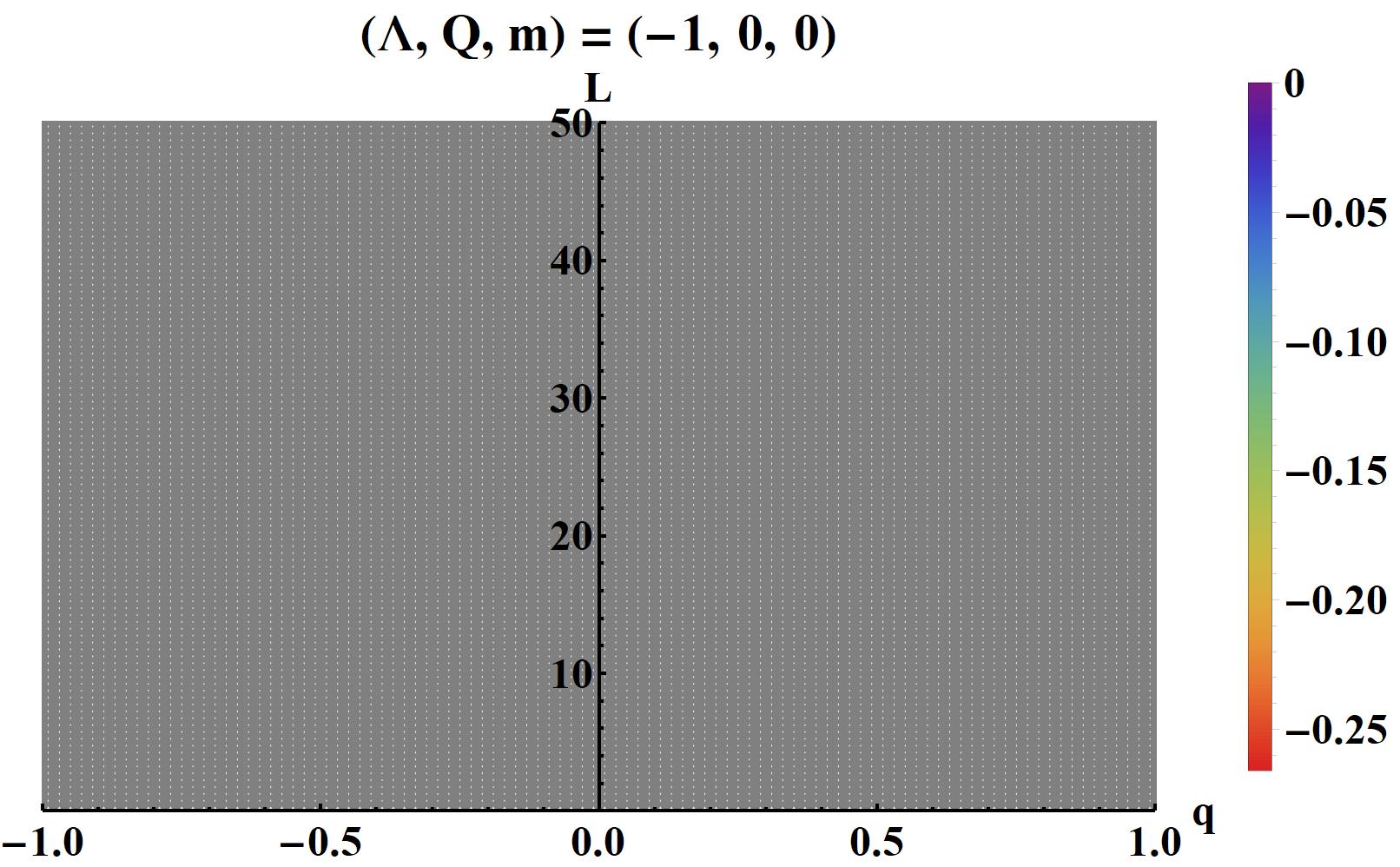}
		\end{minipage} \\

		\begin{minipage}[t]{0.32\hsize}
		\centering
			\includegraphics[width=46mm]{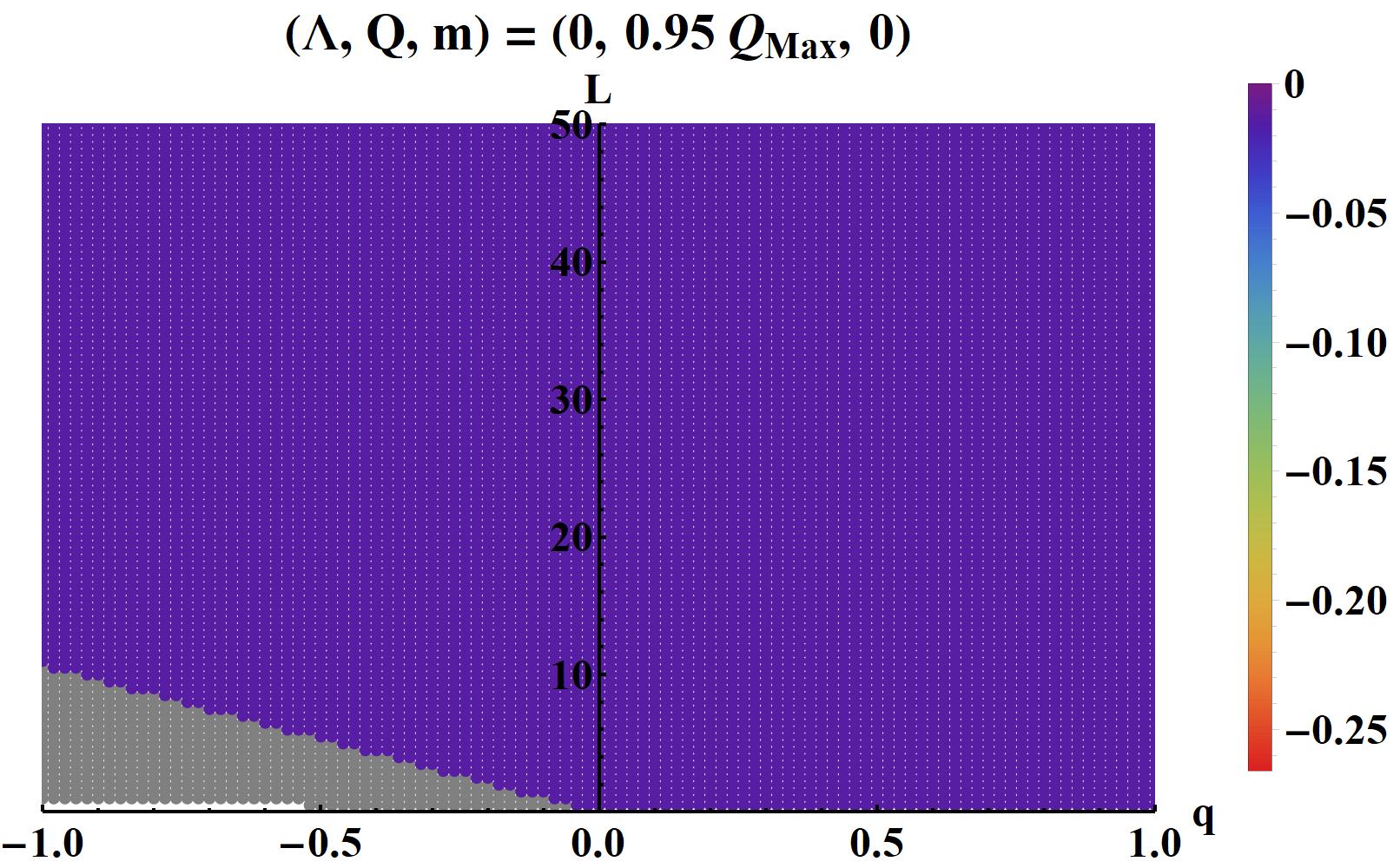}
		\end{minipage} &
		\begin{minipage}[t]{0.32\hsize}
		\centering
			\includegraphics[width=46mm]{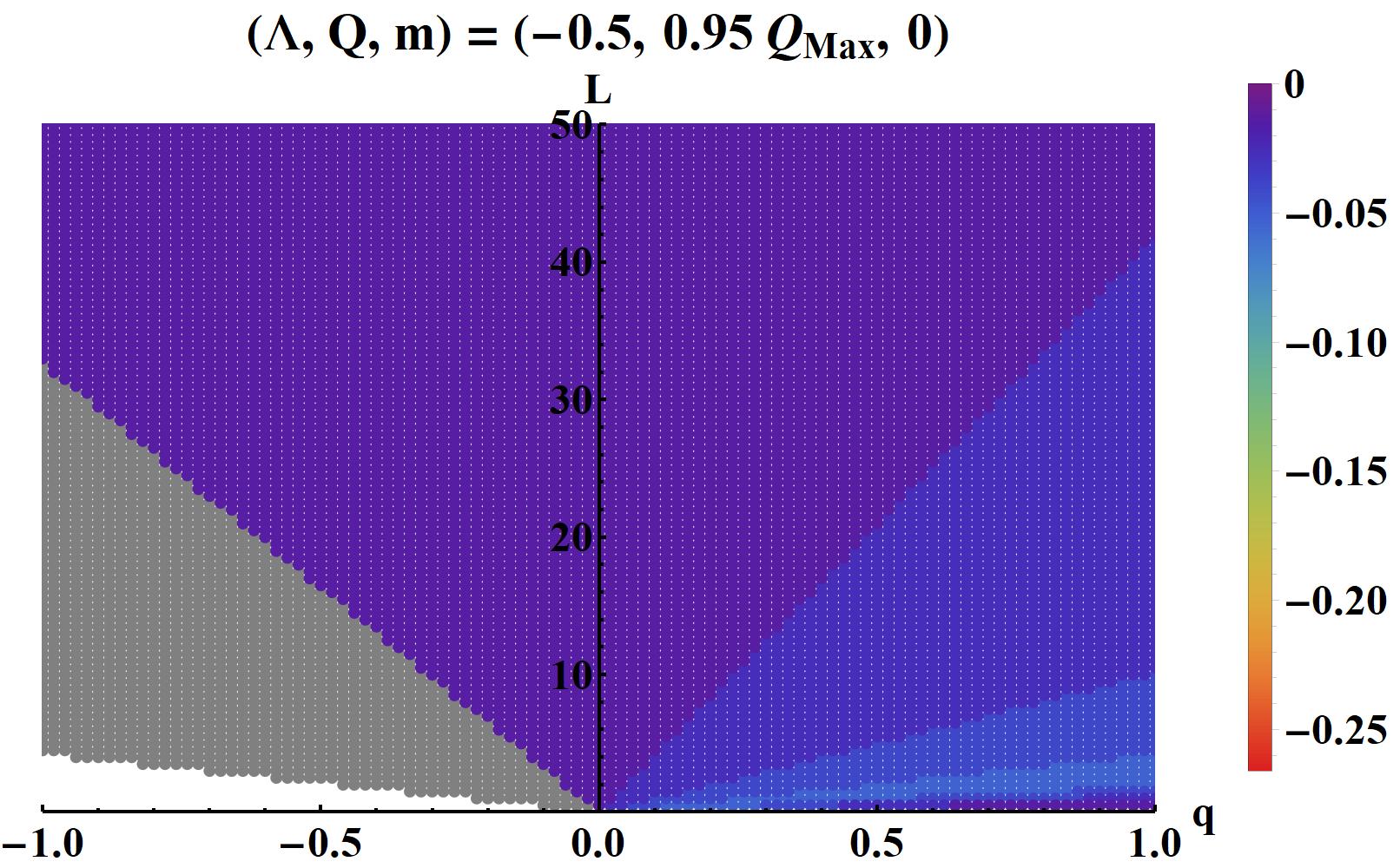}
		\end{minipage} &
		\begin{minipage}[t]{0.32\hsize}
		\centering
			\includegraphics[width=46mm]{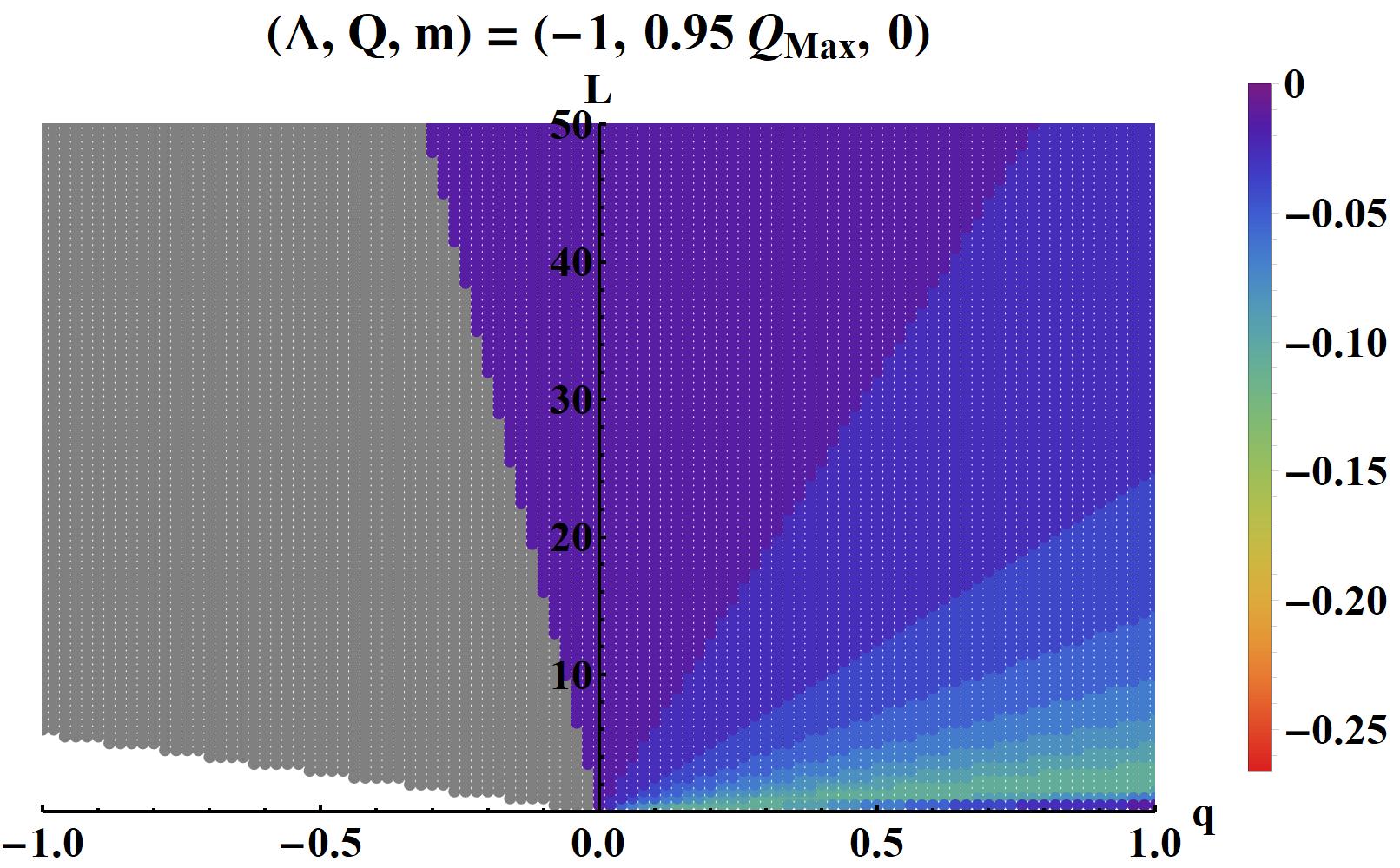}
		\end{minipage} \\
		
		\begin{minipage}[t]{0.32\hsize}
		\centering
			\includegraphics[width=46mm]{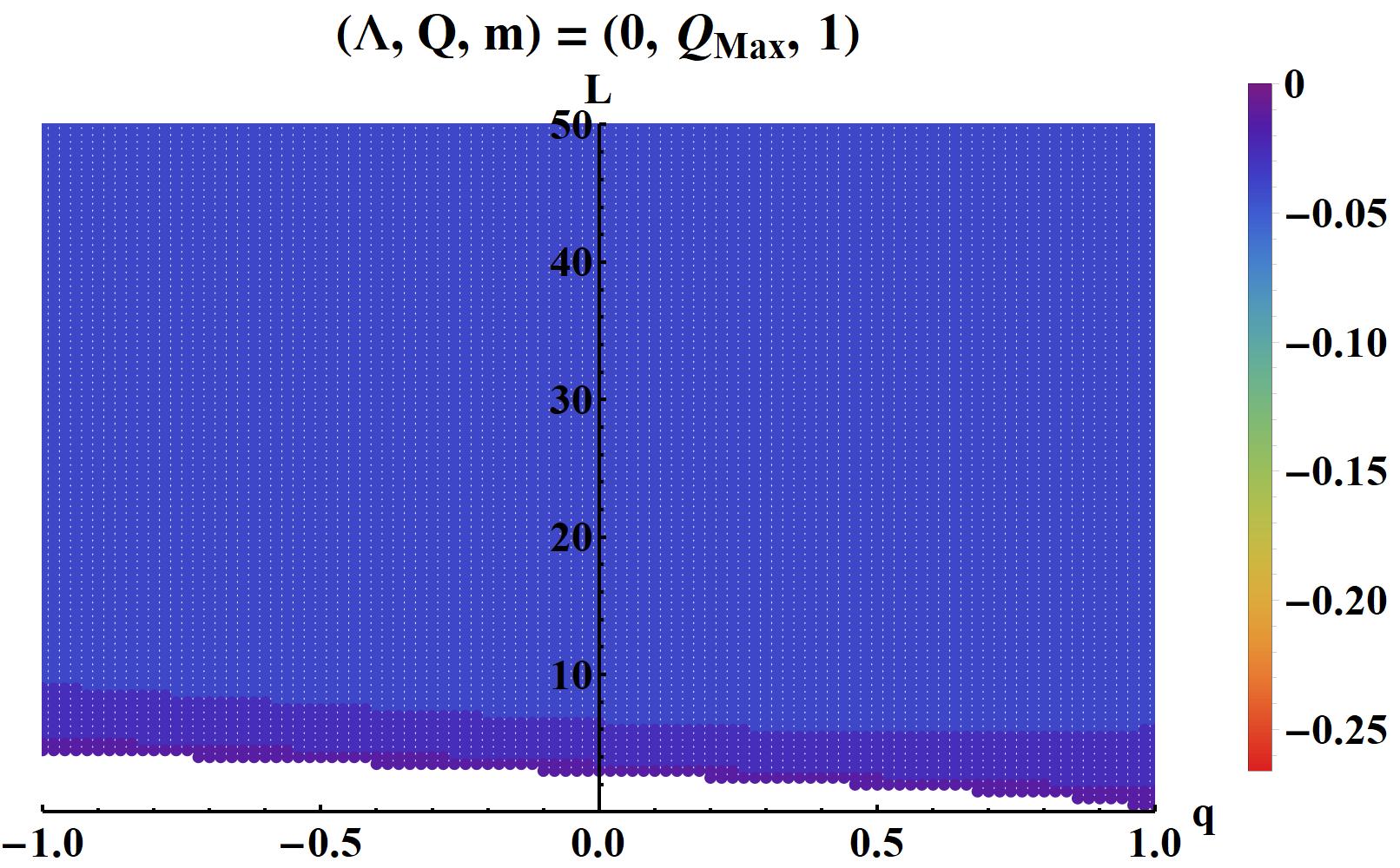}
		\end{minipage} &
		\begin{minipage}[t]{0.32\hsize}
		\centering
			\includegraphics[width=46mm]{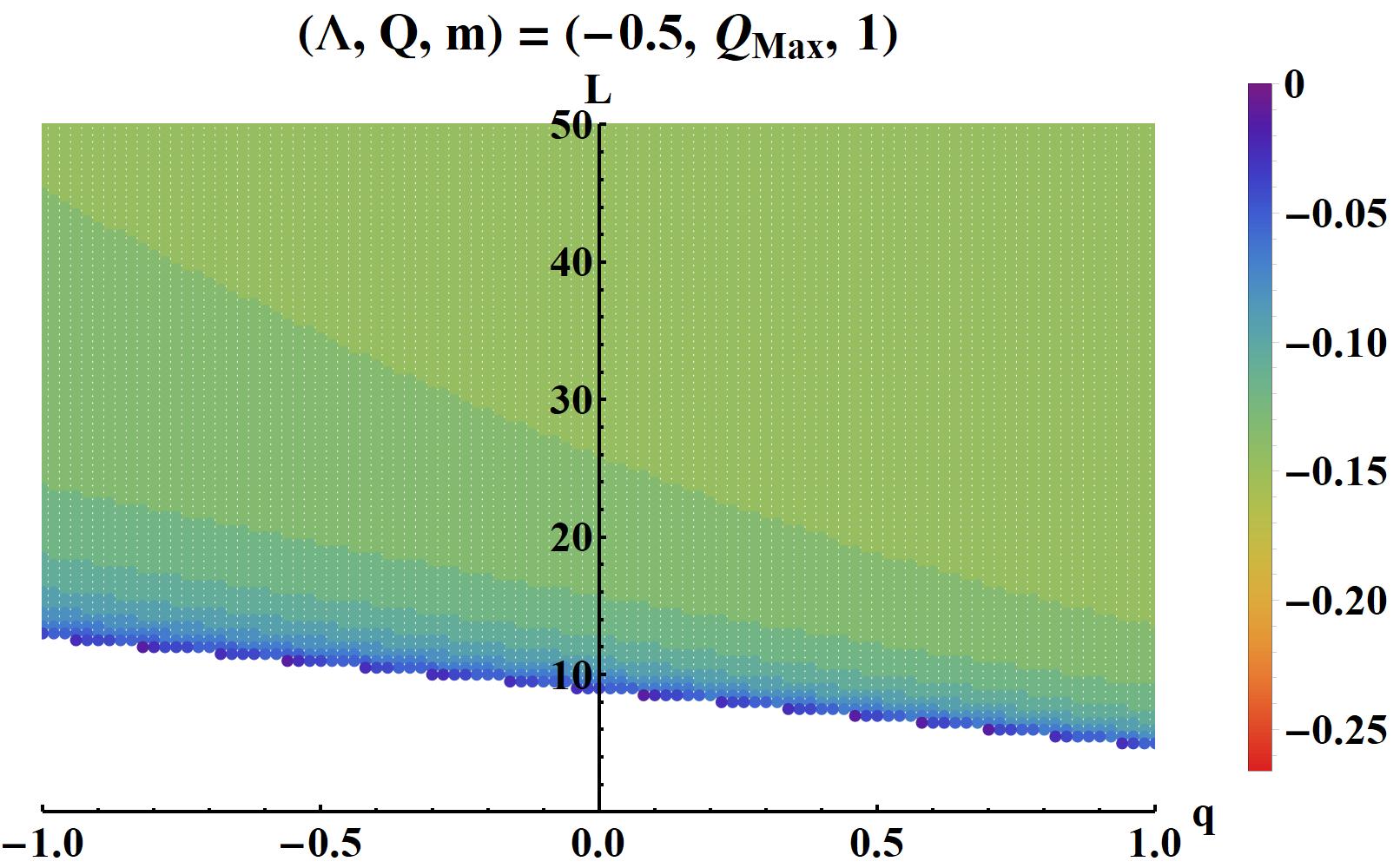}
		\end{minipage} &
		\begin{minipage}[t]{0.32\hsize}
		\centering
			\includegraphics[width=46mm]{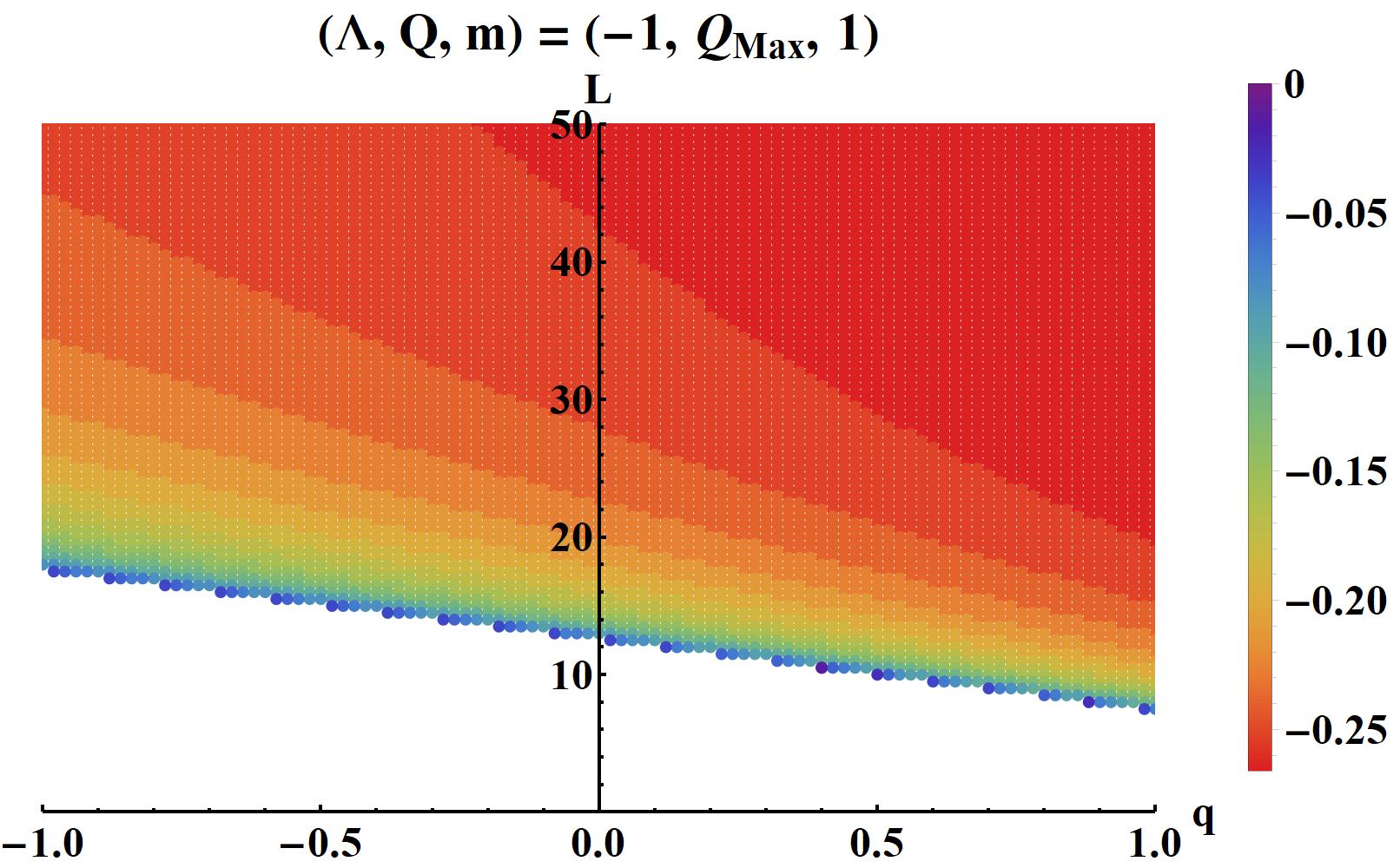}
		\end{minipage} 
	\end{tabular}
	\caption{The result for the particle with $m=0$ in the RN-AdS black hole.\label{fig:m=0_RN}}
\end{figure}
\endgroup

\begingroup
\renewcommand{\arraystretch}{5}
\begin{figure}[htbp]
	\begin{tabular}{ccc}
		\begin{minipage}[t]{0.32\hsize}
		\centering
			\includegraphics[width=46mm]{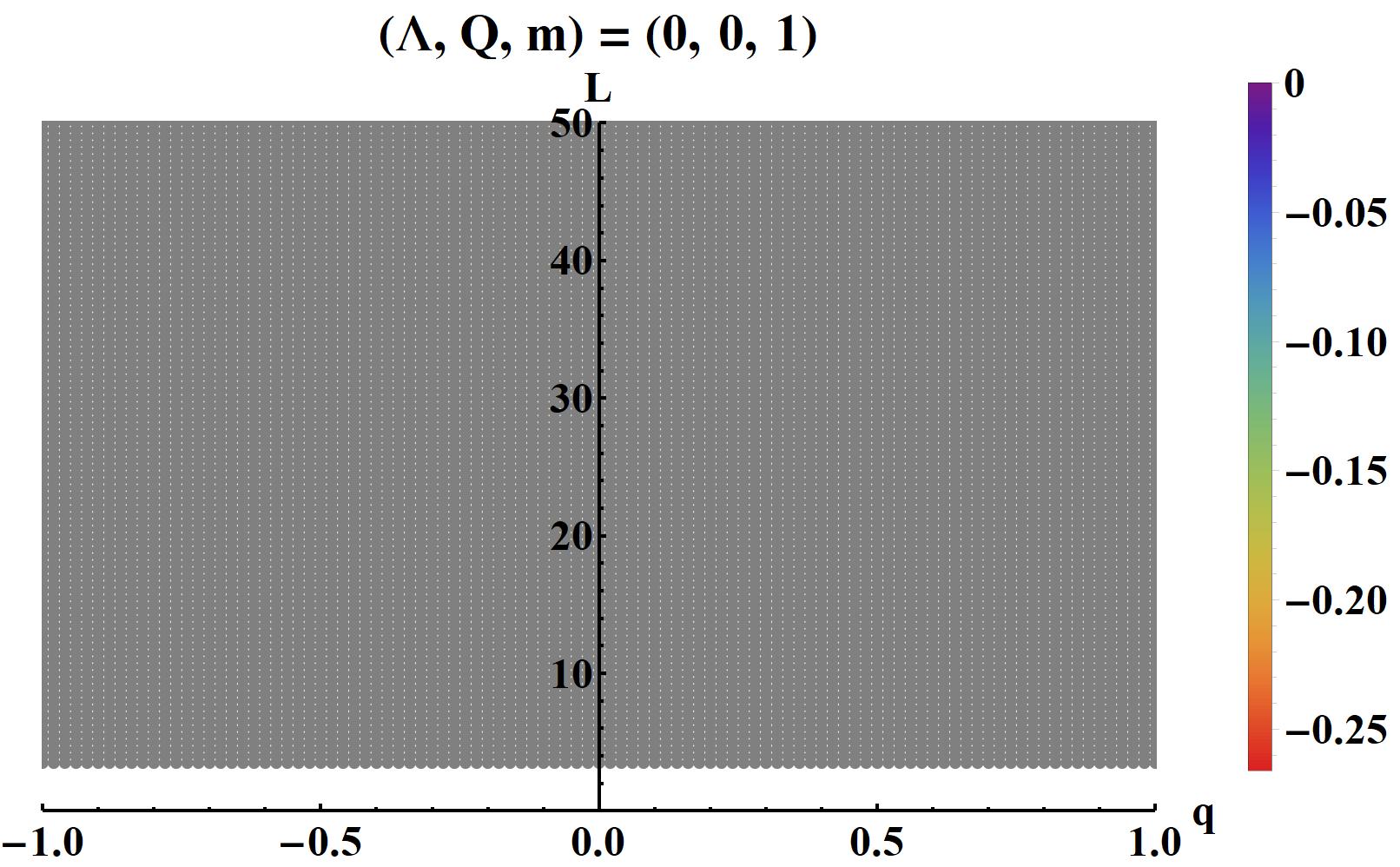}
		\end{minipage} &
		\begin{minipage}[t]{0.32\hsize}
		\centering
			\includegraphics[width=46mm]{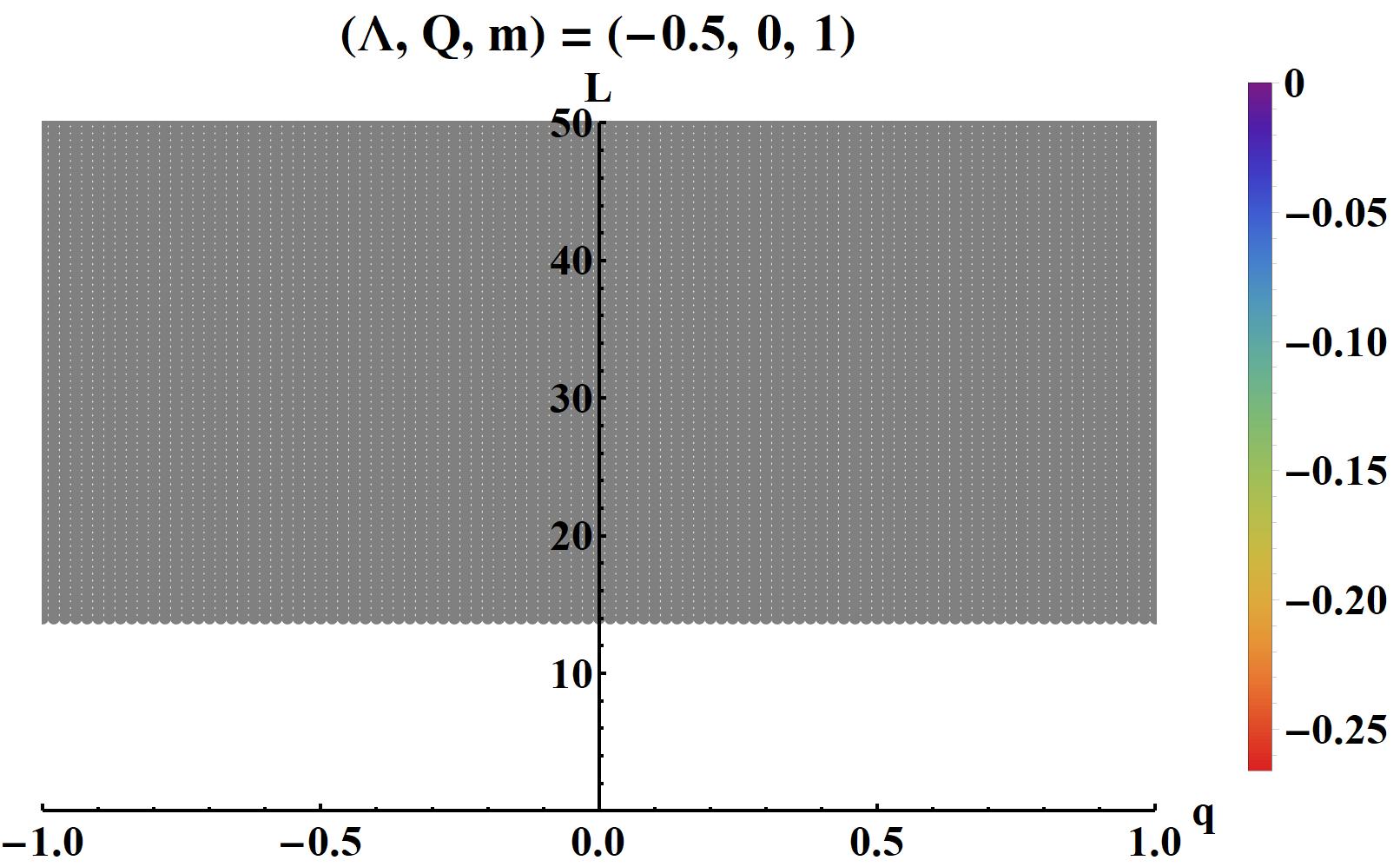}
		\end{minipage} &
		\begin{minipage}[t]{0.32\hsize}
		\centering
			\includegraphics[width=46mm]{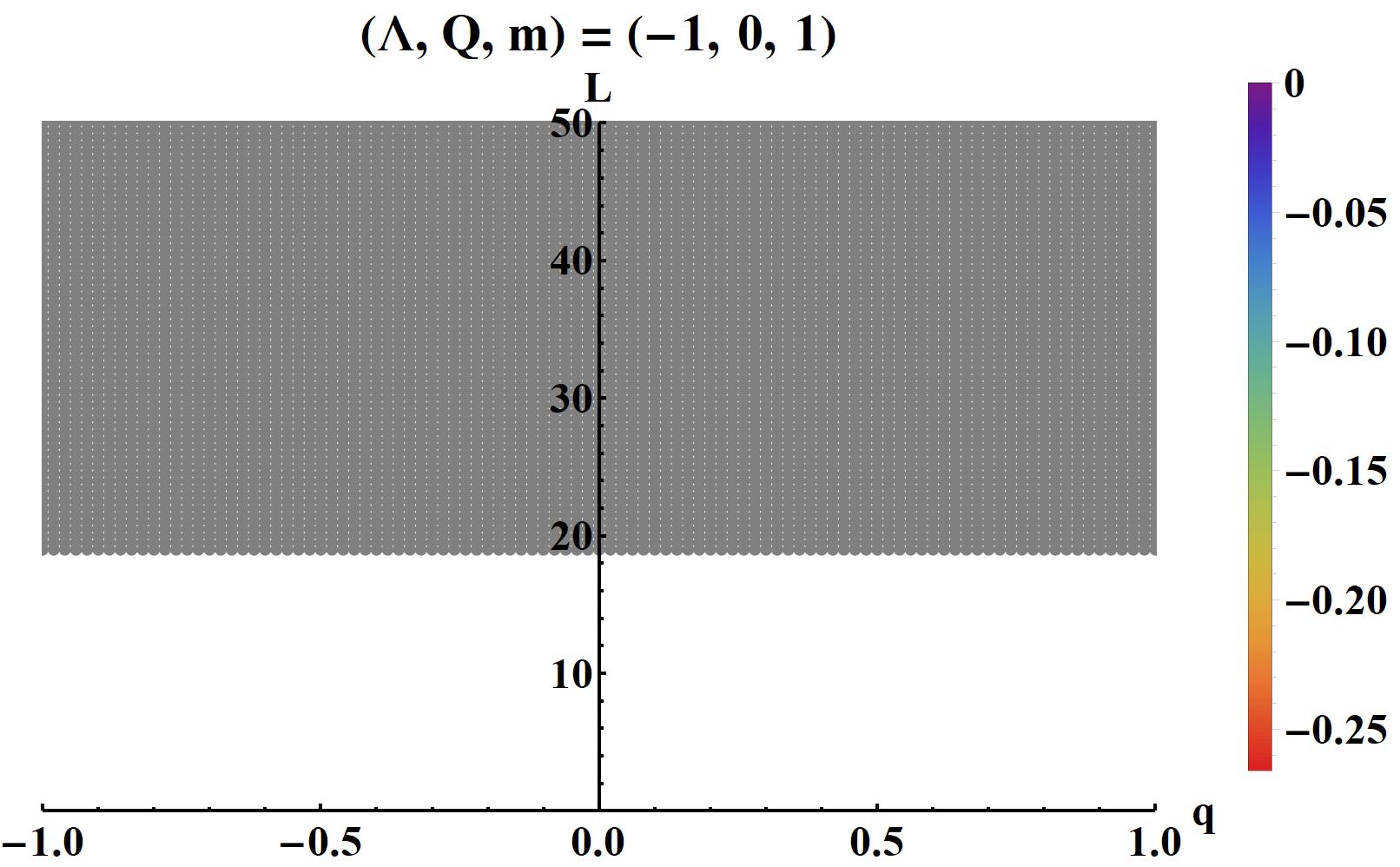}
		\end{minipage} \\

		\begin{minipage}[t]{0.32\hsize}
		\centering
			\includegraphics[width=46mm]{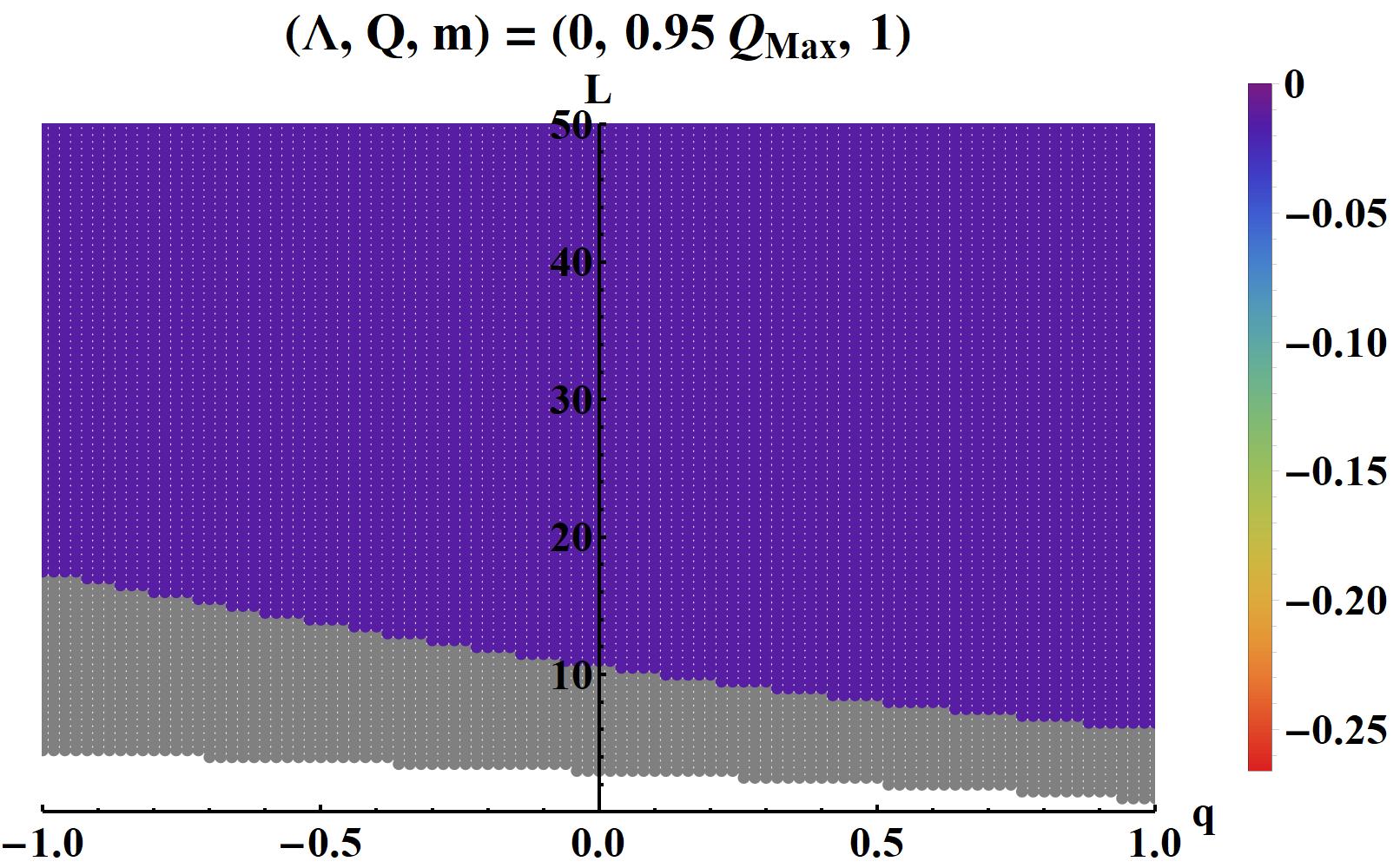}
		\end{minipage} &
		\begin{minipage}[t]{0.32\hsize}
		\centering
			\includegraphics[width=46mm]{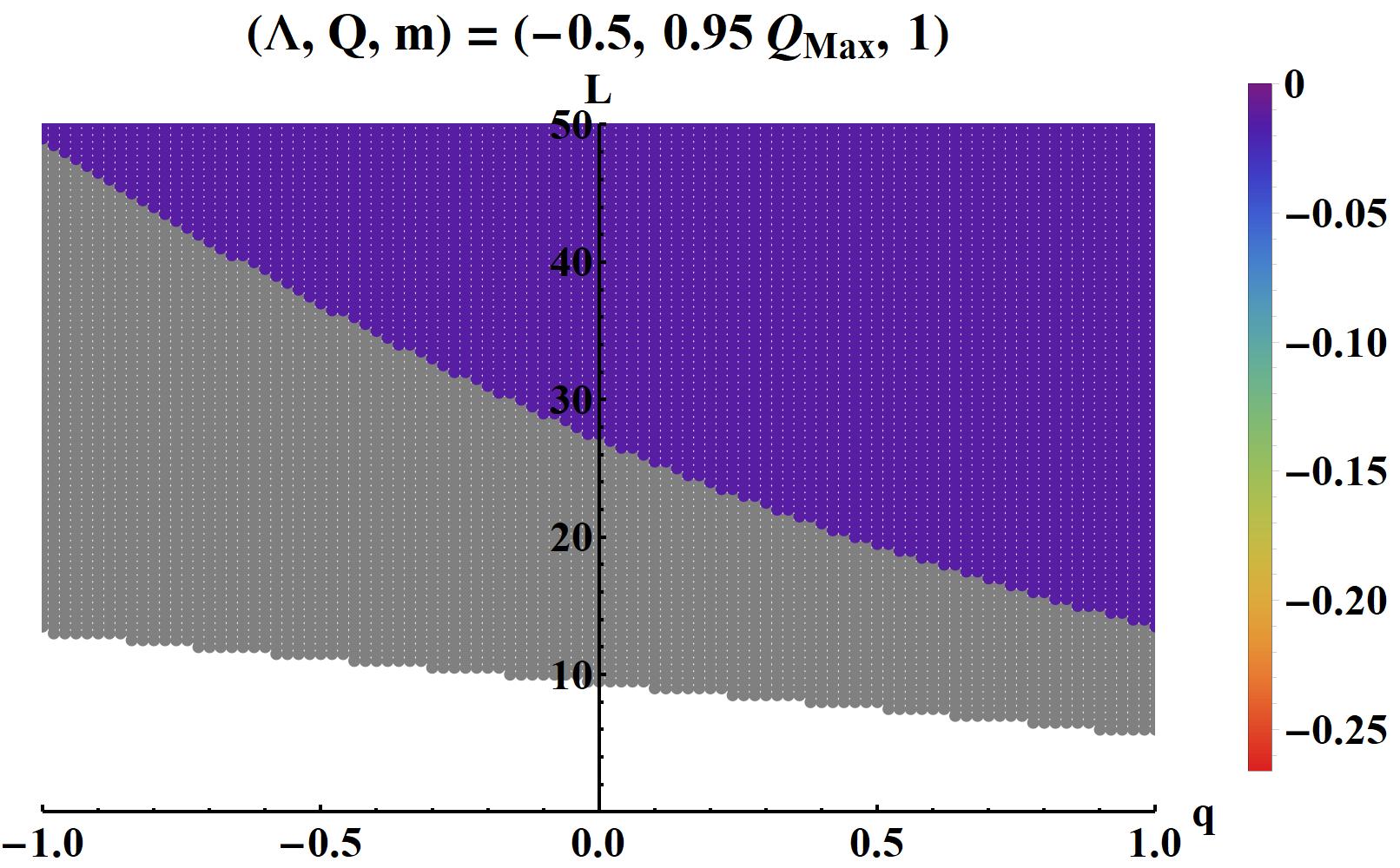}
		\end{minipage} &
		\begin{minipage}[t]{0.32\hsize}
		\centering
			\includegraphics[width=46mm]{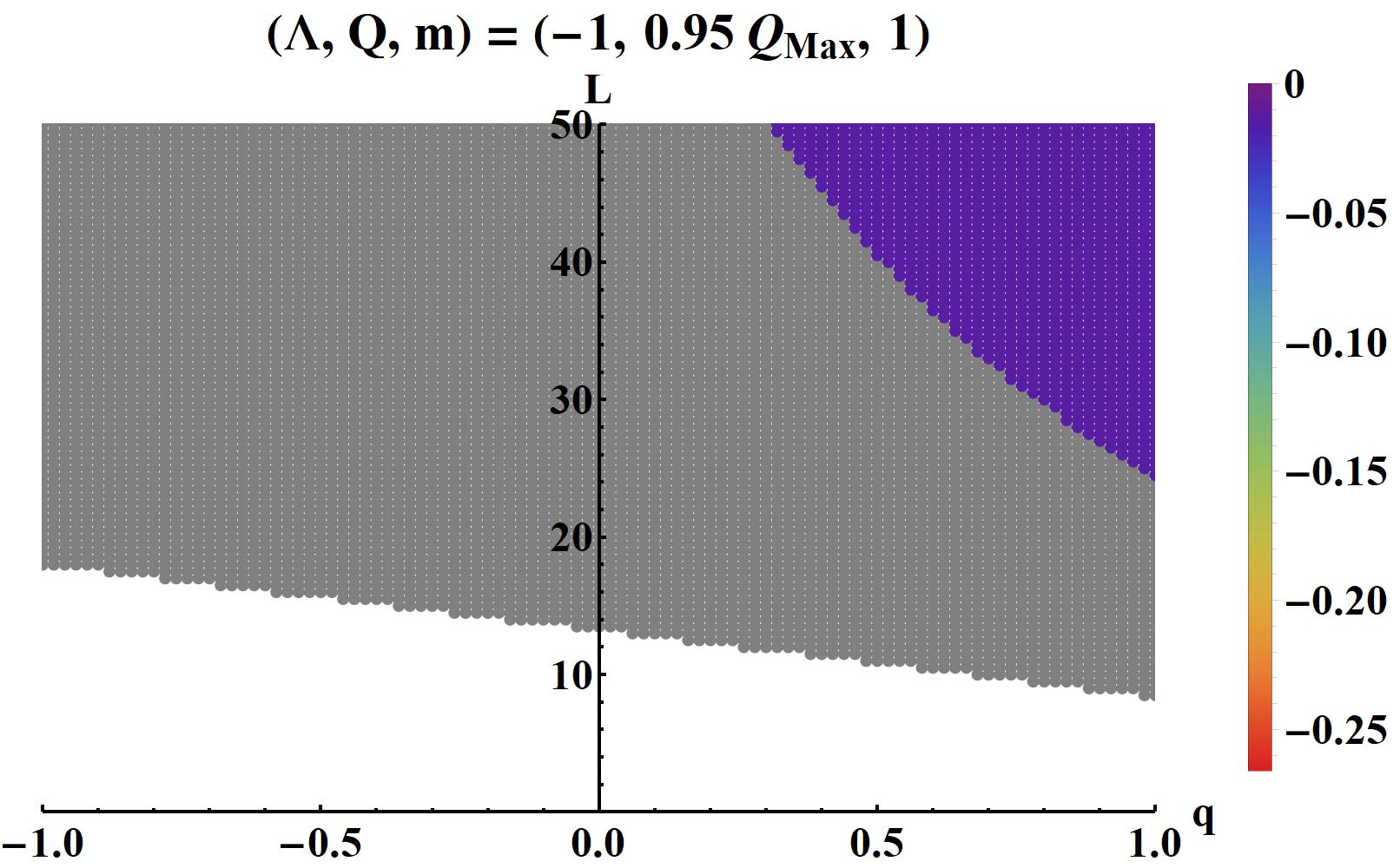}
		\end{minipage} \\
		
		\begin{minipage}[t]{0.32\hsize}
		\centering
			\includegraphics[width=46mm]{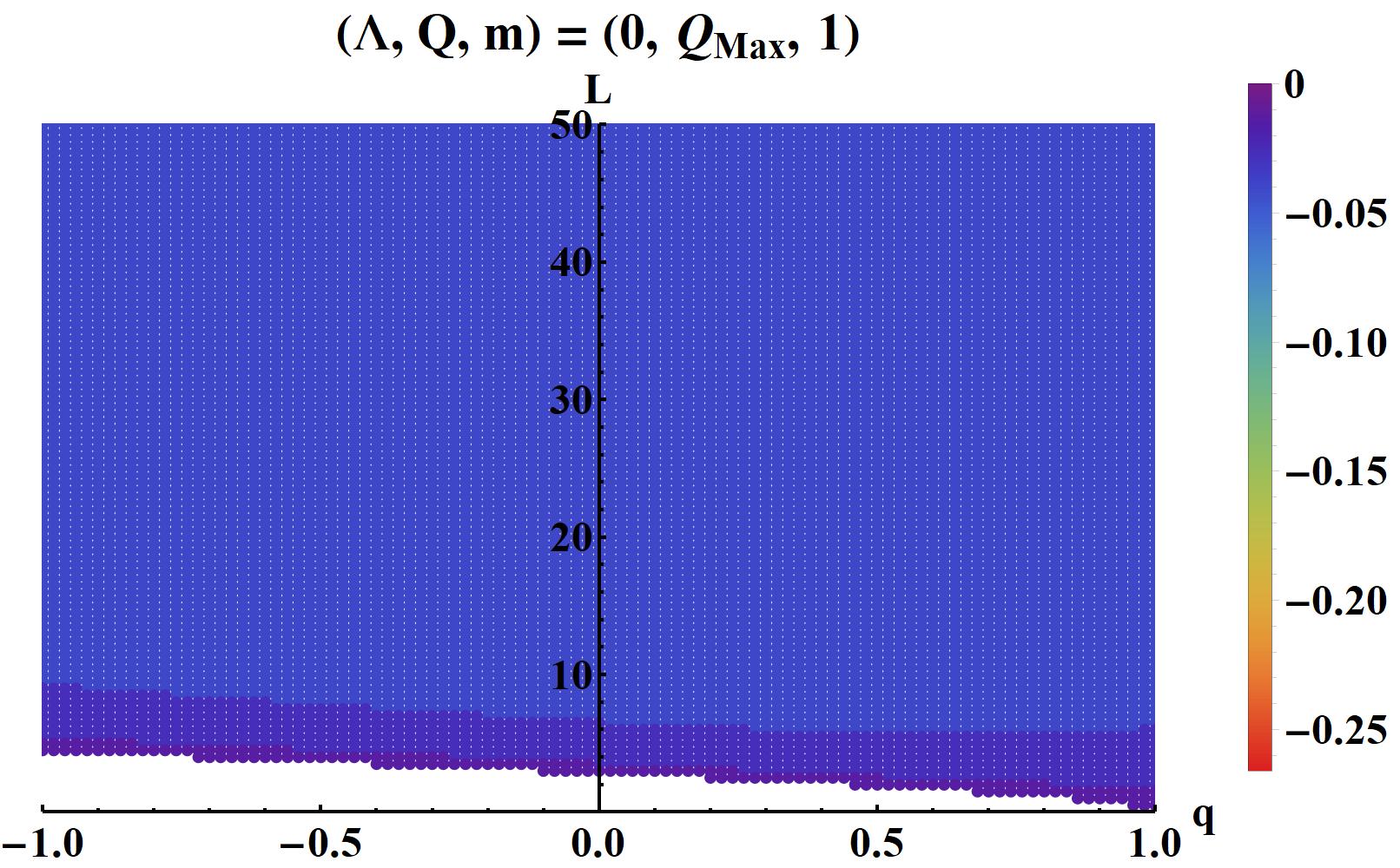}
		\end{minipage} &
		\begin{minipage}[t]{0.32\hsize}
		\centering
			\includegraphics[width=46mm]{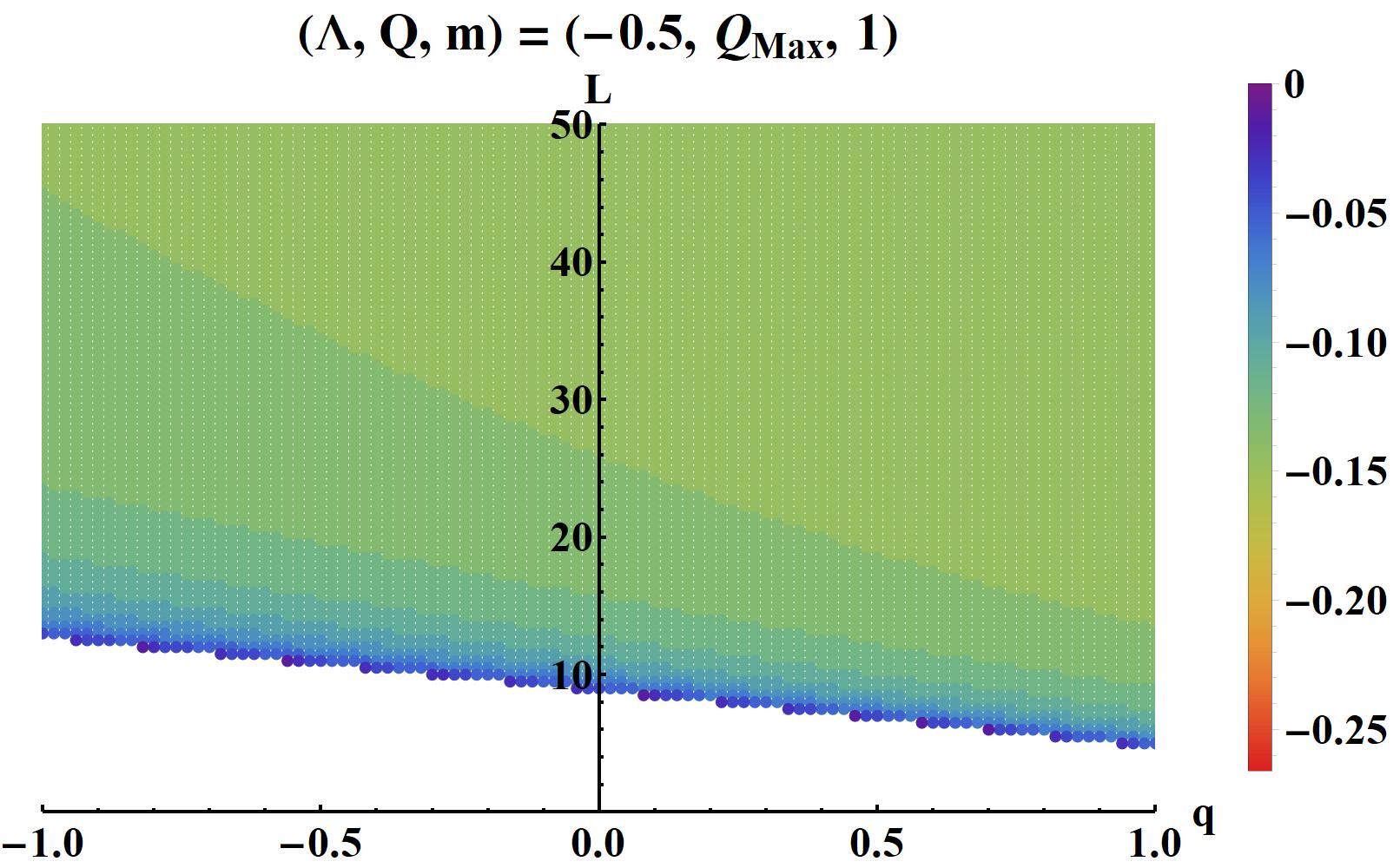}
		\end{minipage} &
		\begin{minipage}[t]{0.32\hsize}
		\centering
			\includegraphics[width=46mm]{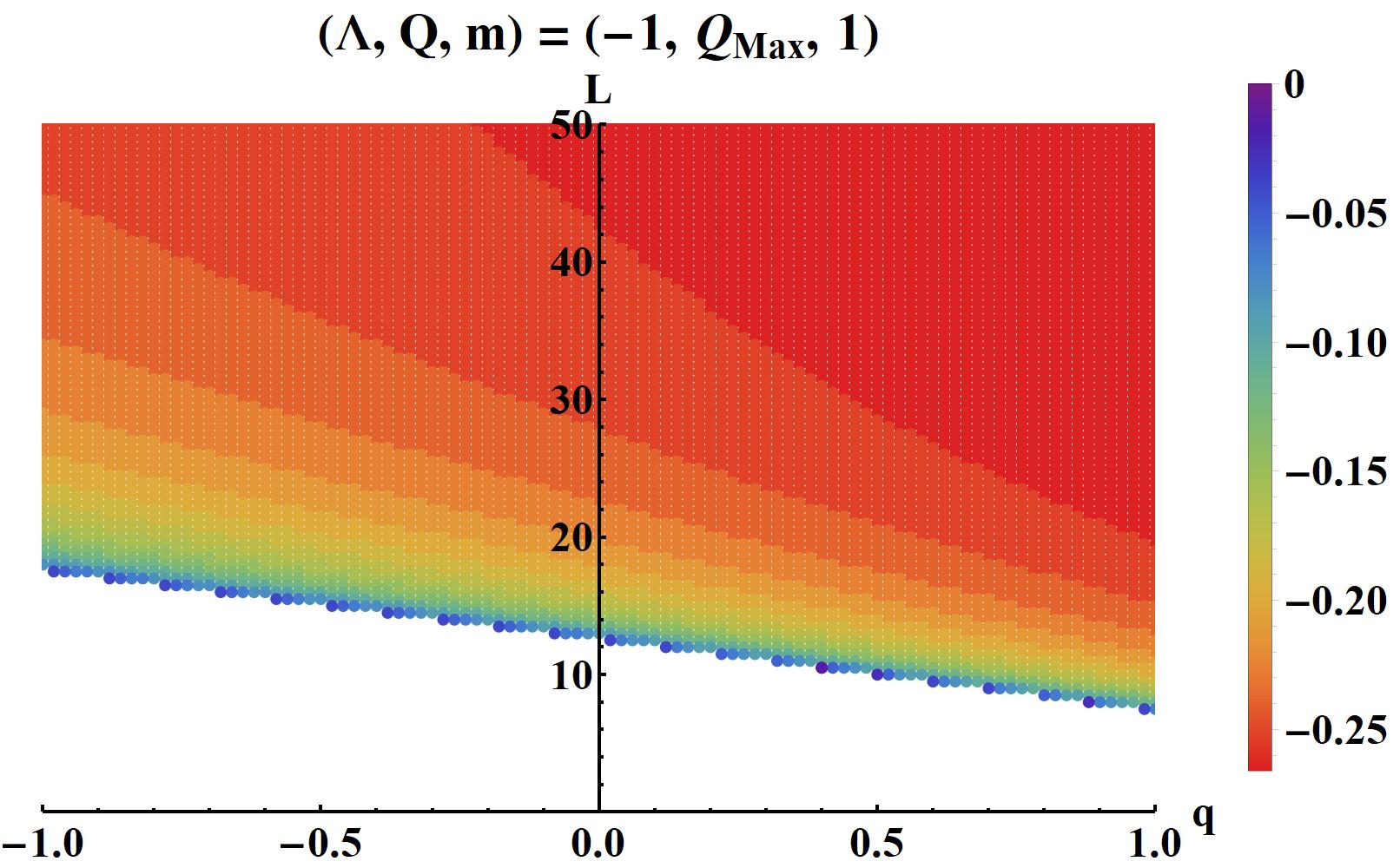}
		\end{minipage} 
	\end{tabular}
	\caption{The result for the particle with $m=1$ in the RN-AdS black hole. \label{fig:m=1_RN}}
\end{figure}
\endgroup

\subsection{The limits of the large negative cosmological constant and near horizon}
So far, we have numerically investigated the bound on the Lyapunov exponent for the RN-AdS black hole.
In this subsection, we consider three limits of the RN-AdS black hole: the large negative cosmological constant and near-horizon.

Substituting $a=0$ for \eqref{eq:eff_potential}, we obtain the effective potential for the RN-AdS black hole,
	\begin{align}
		V_{\rm eff}(r)=\frac{\sqrt{\Delta_r\left(L^2+m^2 r^2\right)}+q Q r}{r^2}.
	\end{align}
We can solve $V_{\rm eff}'(r)=0$ for the particle charge $q$:
	\begin{align}
		q=\left.\frac{L^2 r \Delta_r'-4 L^2\Delta_r+m^2 r^3 \Delta_r'-2 m^2 r^2 \Delta_r}{2 Q r \sqrt{\Delta_r\left(L^2+m^2 r^2\right)}}\right|_{r=r_0},
	\label{eq:qSol_RN_AdS}
	\end{align}
where $r_0$ is a position of local extremum.
Then, $r_0(>r_+)$ is a parameter of the Lyapunov exponent instead of $q$.
The Lyapunov exponent is then evaluated by using \eqref{eq:qSol_RN_AdS}.

Let us investigate the bound in the case of a large negative cosmological constant.
The leading term of $\kappa^2-\lambda^2$ for a large negative $\Lambda$ is given by
	\begin{align}
		\kappa^2-\lambda^2=&\frac{(r_0-r_+)\Lambda^2}{36 r_0^4 \left(L^2+m^2 r_0^2\right)^2}\bigg(3 L^4 r_+^2 (r_0-r_+) \left(3 r_0^2+2 r_+ r_0+r_+^2\right) \nonumber \\
		&\qquad\qquad\qquad\qquad\quad+m^4 r_0^4 \left(8 r_0^5+8r_+ r_0^4 +17 r_+^2 r_0^3+r_+^3r_0^2+r_+^4r_0+r_+^5\right) \nonumber \\
		&\qquad\qquad\qquad\qquad\quad+6 L^2 m^2 r_0^2 \left(2 r_0^5+2 r_+r_0^4+5r_+^2 r_0^3-r_+^3r_0^2-r_+^4r_0-r_+^5\right)\bigg)+{\cal O}(\Lambda).
	\end{align}
We can easily show that the leading term on the right-hand side is always positive for $r_0>r_+$.
Thus for a large cosmological constant, the bound is always satisfied, $\kappa^2-\lambda^2>0$.

The other interesting limit is the near horizon limit, $r_0\to r_+$.
To examine this limit, we introduce an infinitesimal parameter $\epsilon$ as $r_0=r_++\epsilon$.
We then expand $\kappa^2-\lambda^2$ around $\epsilon=0$:
	\begin{align}
		\kappa^2-\lambda^2=\frac{m^2  \left(Q^2+\Lambda  r_+^4-r_+^2\right)^2}{r_+^5 \left(L^2+m^2 r_+^2\right)}\epsilon+{\cal O}(\epsilon^{2}).
	\end{align}
The leading term on the right-hand side is always positive, thus, the bound is definitely satisfied in the near-horizon limit.

\section{The bound for rotating AdS black holes}
Next, we discuss the role of the angular momentum of the AdS black hole.
As such, we first consider the Kerr-AdS black hole and combine the rotating effect with the charge effect, namely, we consider the Kerr-Newman-AdS black hole.

\subsection{The Kerr-AdS black hole}
We show the numerical results of $\kappa^2-\lambda^2$ in Fig. \ref{fig:m=0_Kerr} and \ref{fig:m=1_Kerr}.
We suppose $M=1$.
In the Kerr-AdS black hole, the charge of the particle $q$ is no longer relevant since there are no electromagnetic interactions between the particle and the black hole.
The horizontal axis represents the angular momentum of the black hole $a$, and the vertical axis represents the cosmological constant for each plot in Fig. \ref{fig:m=0_Kerr} and \ref{fig:m=1_Kerr}.
As in the case of RN-AdS black hole, the colored, gray, and white regions represent the violation, chaotic and non-chaos cases, respectively.
We see that when the cosmological constant is turned on, the violated region becomes smaller.

To investigate the effect of the angular momentum of the black hole, we focus on the middle plot in Fig. \ref{fig:m=0_Kerr} and \ref{fig:m=1_Kerr}.
The plot is point symmetric, and we examine the negative $L$ region.
For a small $a$, we are in the gray region; the particle is chaotic, but the bound is satisfied.
The bound is violated only when $a$ is large.
In addition, comparing the result for the massless particle (Fig. \ref{fig:m=0_Kerr}) with that obtained for the massive particle (Fig. \ref{fig:m=1_Kerr}), the white region appears in the latter.
This implies that the mass of the particle contributes to a reduction of chaotic behavior.
The same behavior is evident in the other plots.

Interesting behavior appears in the near-extremal region for $m=0$ (Fig. \ref{fig:m=0_Kerr}).
The extremal limits are given by $a=\pm 1.00, \pm 0.887,\pm 0.847$ for $\Lambda=0,0.6,-1$, respectively.
In the near-extremal region, we see that as the cosmological constant becomes larger, the color becomes red.
This implies that $\kappa^2-\lambda^2$ becomes larger, and the bound is more strongly violated.

\begingroup
\renewcommand{\arraystretch}{5}
\begin{figure}[htbp]
	\begin{tabular}{ccc}
		\begin{minipage}[t]{0.32\hsize}
		\centering
			\includegraphics[width=46mm]{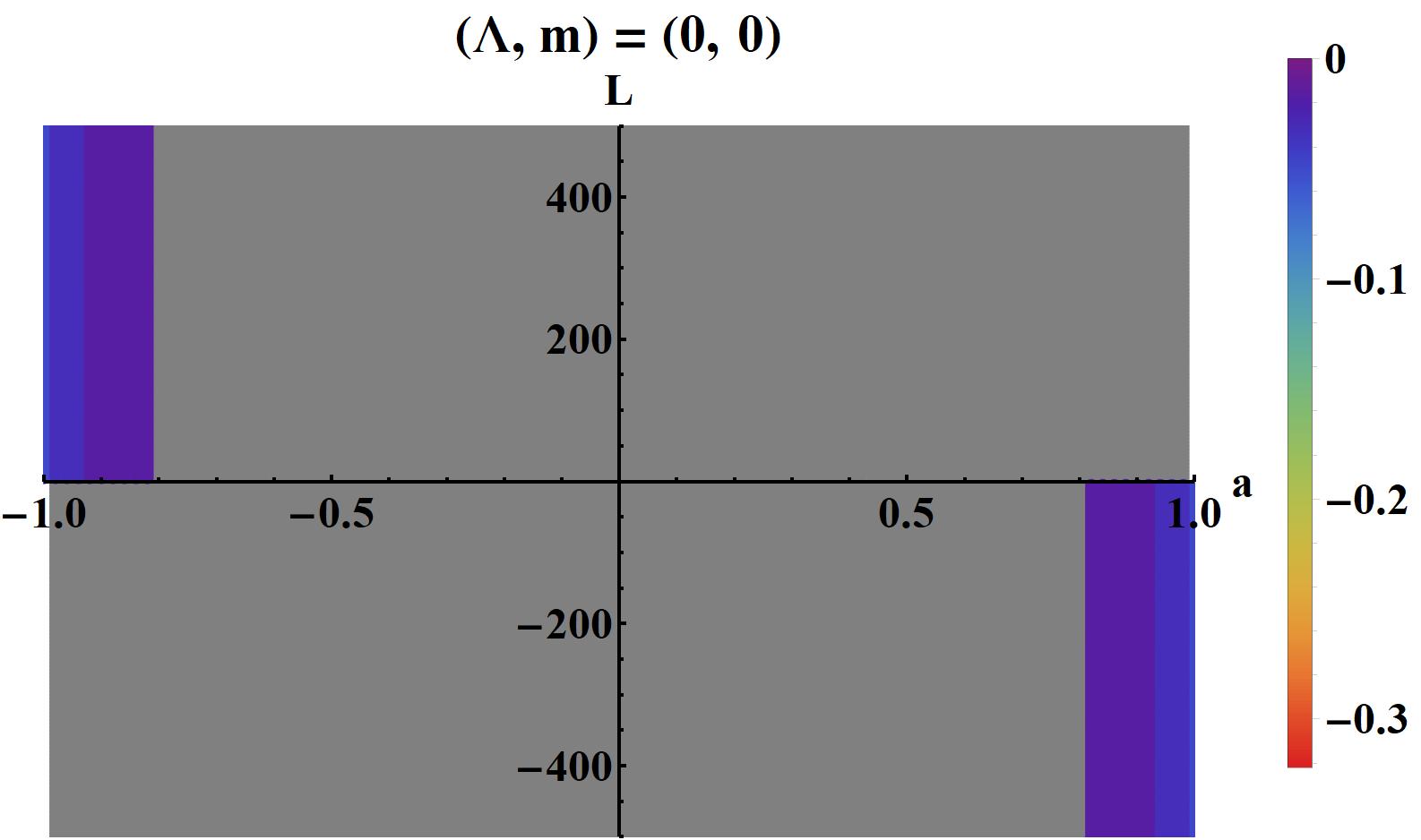}
		\end{minipage} &
		\begin{minipage}[t]{0.32\hsize}
		\centering
			\includegraphics[width=46mm]{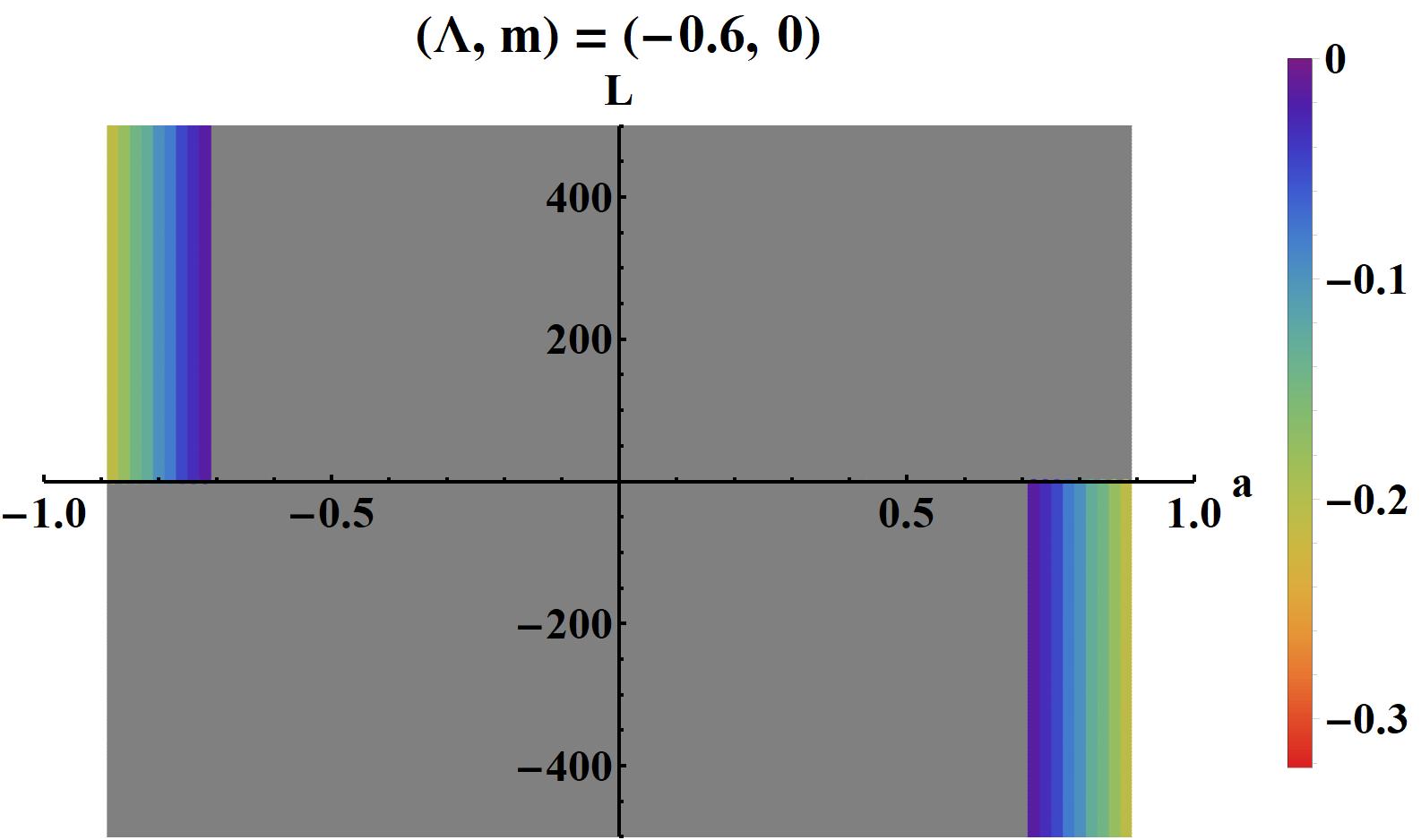}
		\end{minipage} &
		\begin{minipage}[t]{0.32\hsize}
		\centering
			\includegraphics[width=46mm]{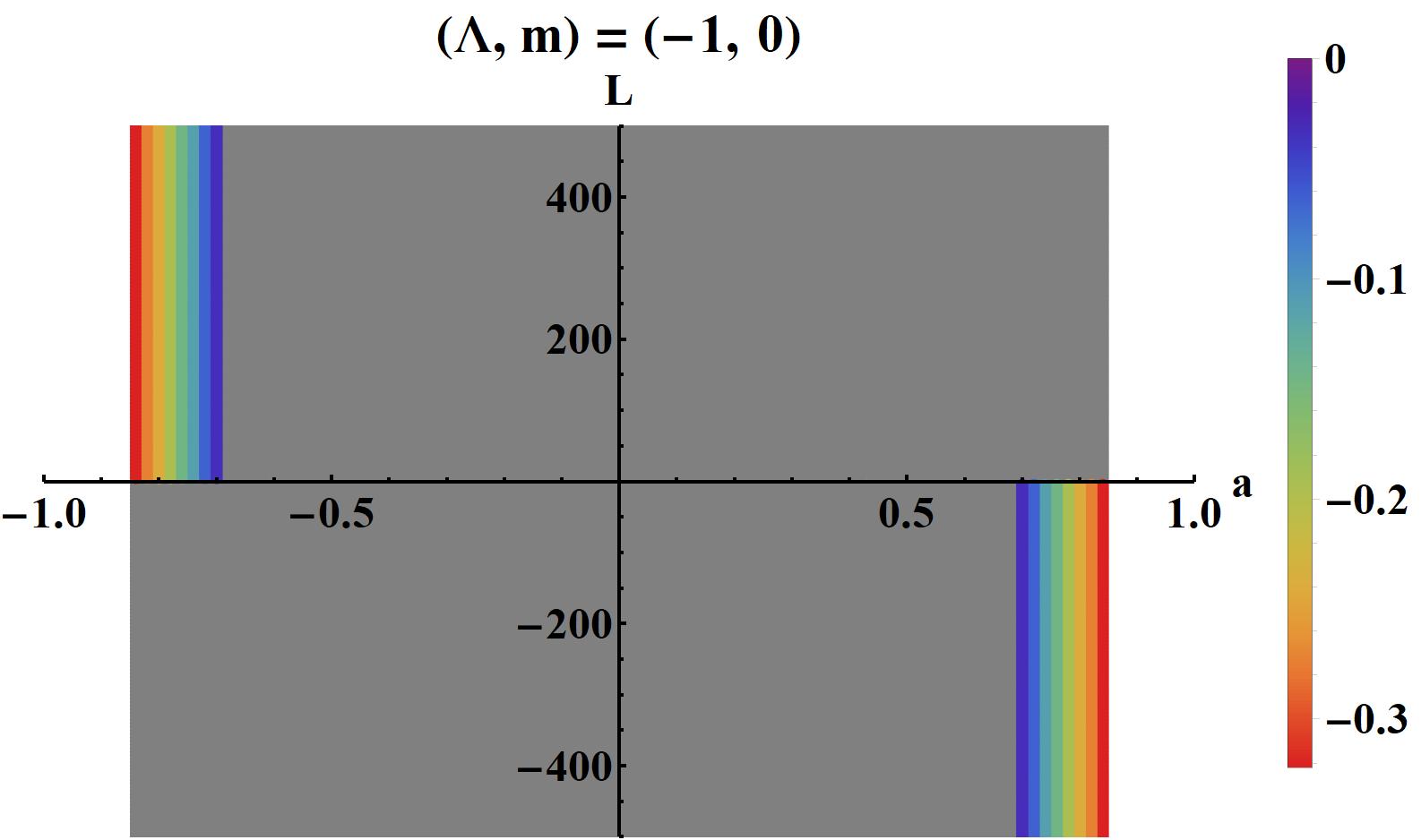}
		\end{minipage} \\
	\end{tabular}
	\caption{The result for the particle with $m=0$ in the Kerr-AdS black hole.\label{fig:m=0_Kerr}}
\end{figure}
\endgroup

\begingroup
\renewcommand{\arraystretch}{5}
\begin{figure}[htbp]
	\begin{tabular}{ccc}
		\begin{minipage}[t]{0.32\hsize}
		\centering
			\includegraphics[width=46mm]{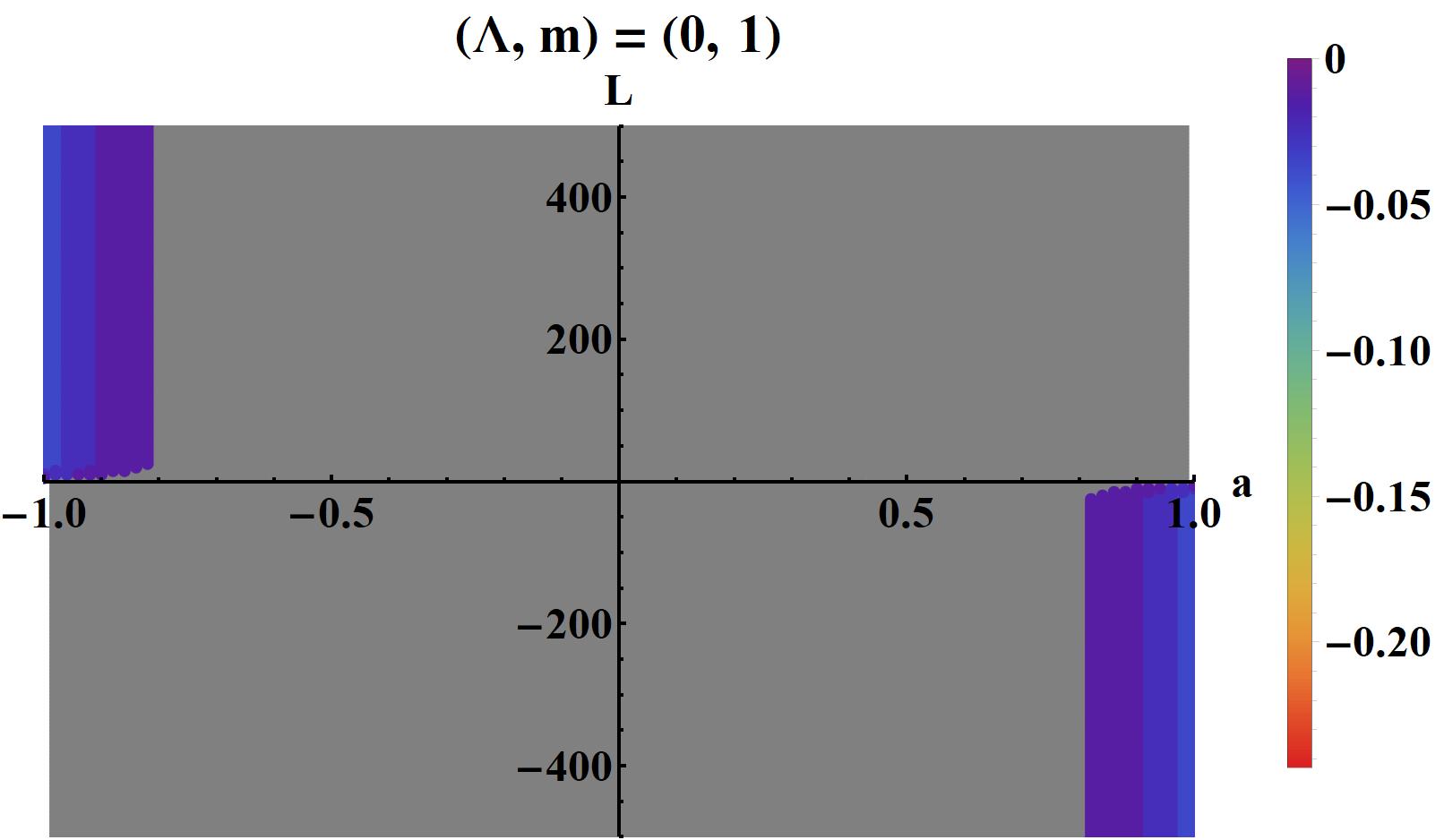}
		\end{minipage} &
		\begin{minipage}[t]{0.32\hsize}
		\centering
			\includegraphics[width=46mm]{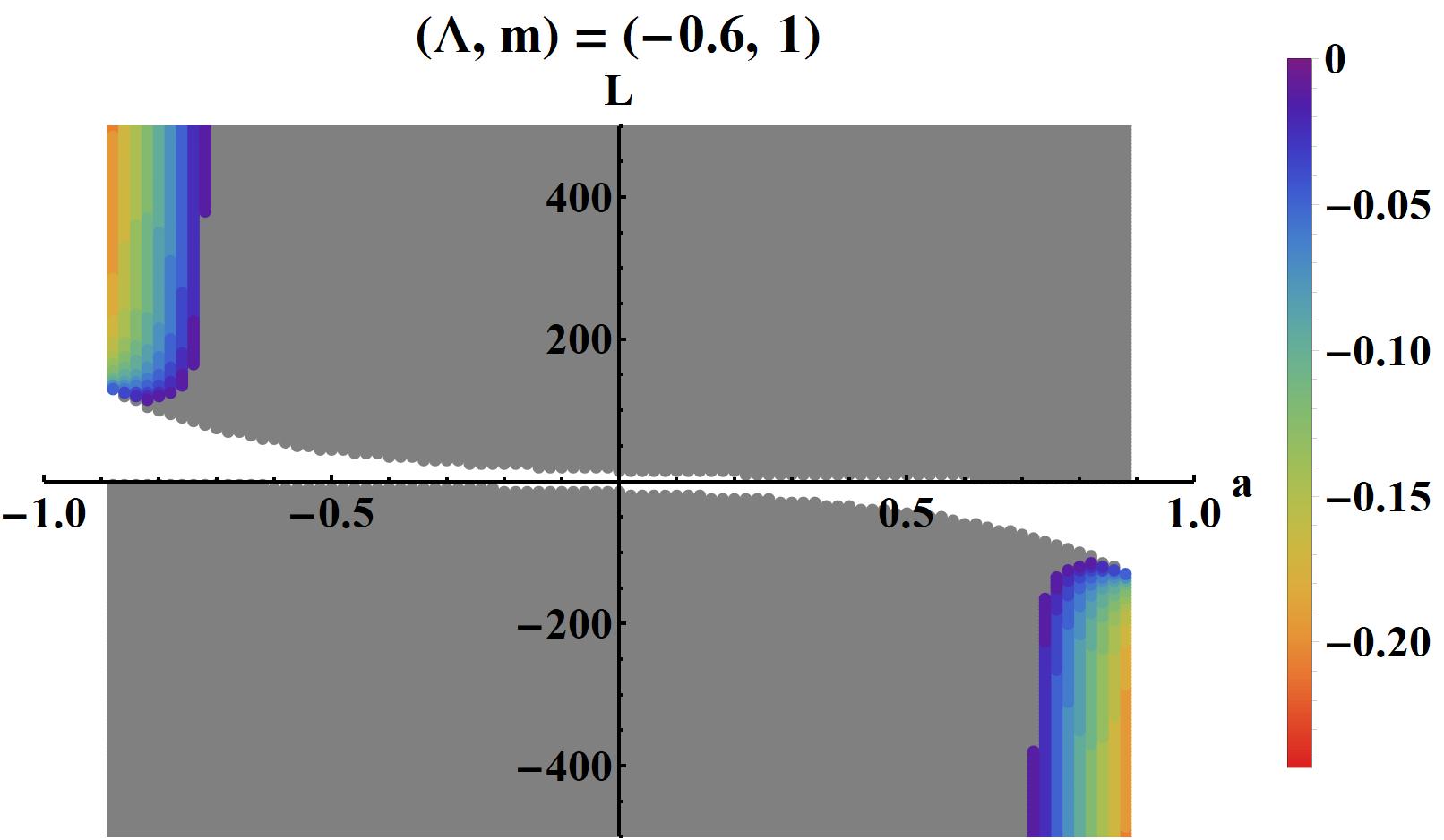}
		\end{minipage} &
		\begin{minipage}[t]{0.32\hsize}
		\centering
			\includegraphics[width=46mm]{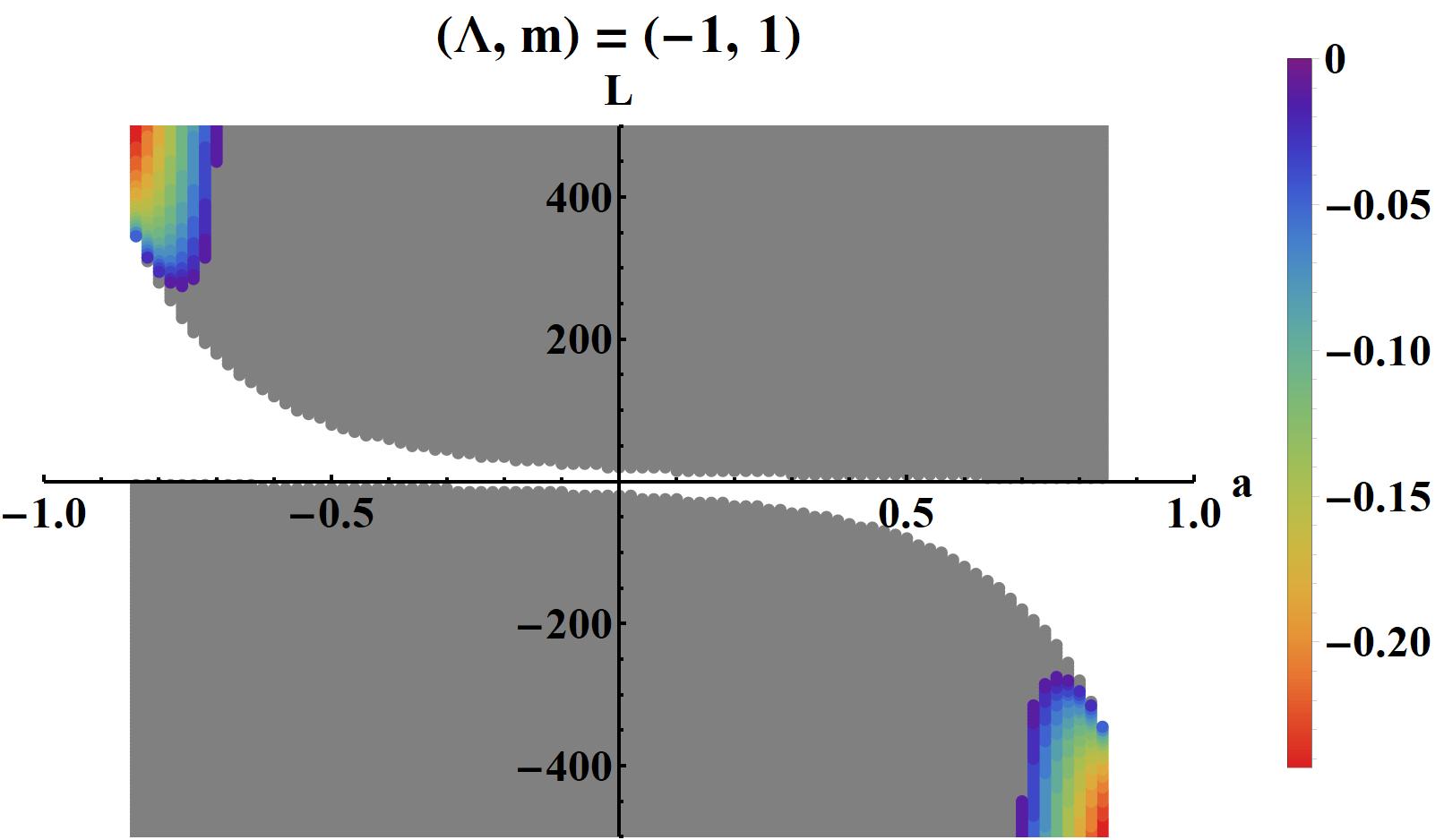}
		\end{minipage} \\
	\end{tabular}
	\caption{The result for the particle with $m=1$ in the Kerr-AdS black hole.\label{fig:m=1_Kerr}}
\end{figure}
\endgroup

\subsection{The Kerr-Newman-AdS black hole}
We show the results of $\kappa^2-\lambda^2$ for the Kerr-Newman black hole in Fig. \ref{fig:m=0_KN} and Fig. \ref{fig:m=1_KN}.
For simplicity, we suppose that the black hole charge and rotation parameters are equivalent and positive, $Q=a>0$.
The extremal black hole has $\Lambda=0, -0.5, -1$ has $Q_{\rm Max}=a_{\rm Max}=0.707, 0.651, 0.619$, respectively.
We arrange nine plots for the massless and massive particles.
The horizontal axis is the charge of the particle $q$, and the vertical axis is the angular momentum of the particle $L$ for each plot.
The three plots in the same row have the same cosmological constant but different black hole charges.
The three plots in the same columns have the same charge and angular momentum of the black hole but different cosmological constants.

Similar to the RN-AdS and Kerr-AdS black holes, we observe that when the cosmological constant is turned on, the colored regions and gray regions become smaller.
We can also observe the effect of the angular momentum of the black hole $a$.
To investigate this aspect, let us compare the first and second plots in the middle column in Fig. \ref{fig:m=0_KN}.
In the first plot, most of the region is gray, whereas, in the second plot, a colored region is observed.
Thus, the angular momentum of the black hole causes the probe particle to be more chaotic.

\begingroup
\renewcommand{\arraystretch}{5}
\begin{figure}[H]
	\begin{tabular}{ccc}
		\begin{minipage}[t]{0.32\hsize}
		\centering
			\includegraphics[width=46mm]{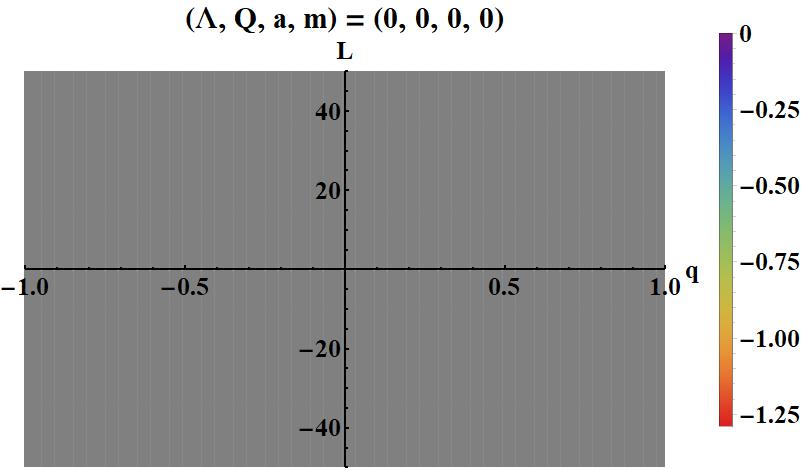}
		\end{minipage} &
		\begin{minipage}[t]{0.32\hsize}
		\centering
			\includegraphics[width=46mm]{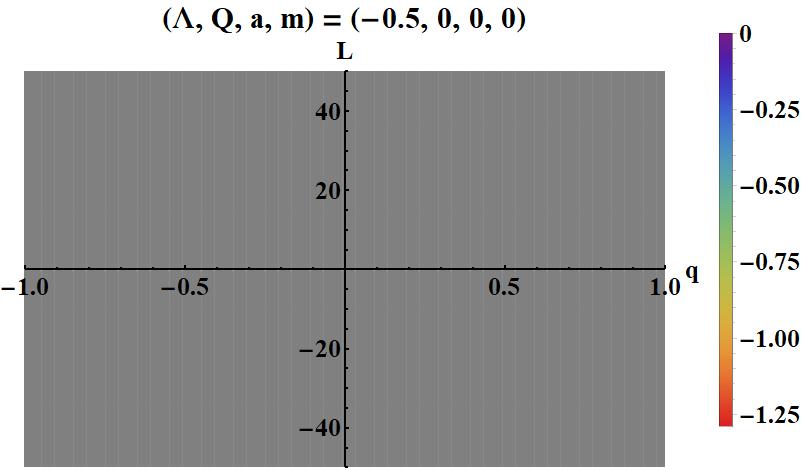}
		\end{minipage} &
		\begin{minipage}[t]{0.32\hsize}
		\centering
			\includegraphics[width=46mm]{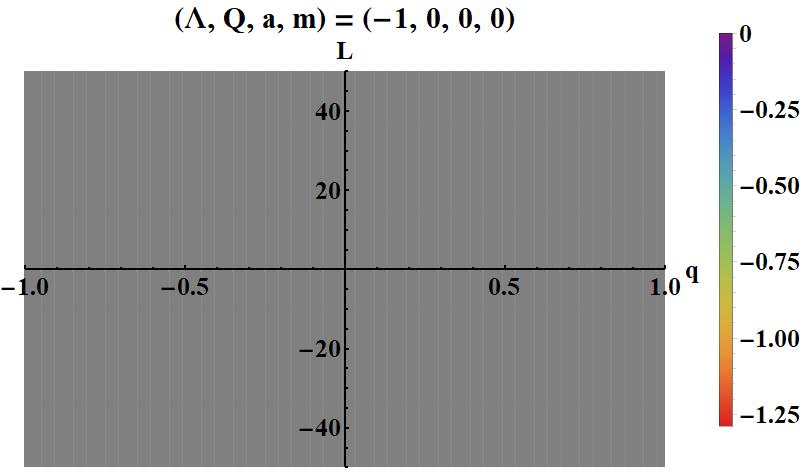}
		\end{minipage} \\

		\begin{minipage}[t]{0.32\hsize}
		\centering
			\includegraphics[width=46mm]{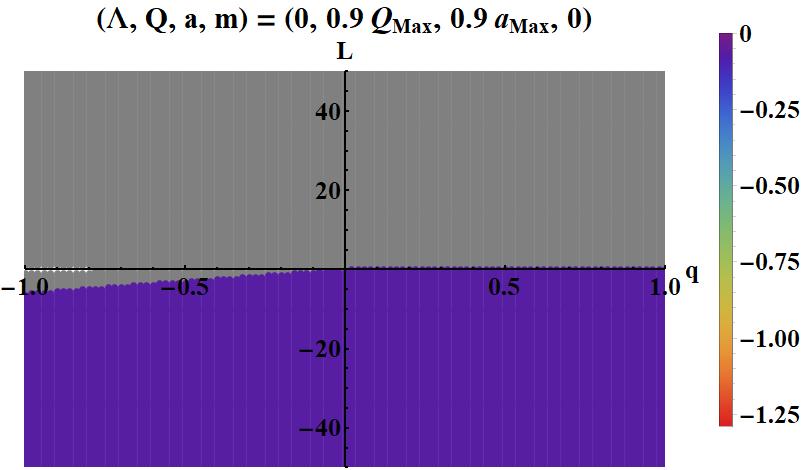}
		\end{minipage} &
		\begin{minipage}[t]{0.32\hsize}
		\centering
			\includegraphics[width=46mm]{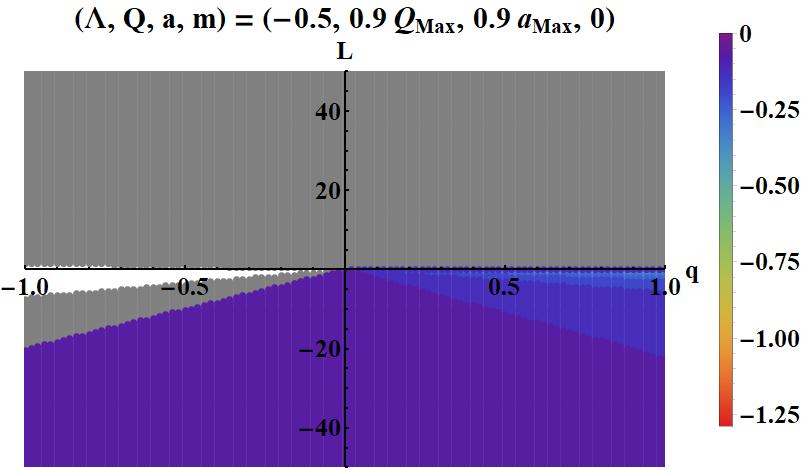}
		\end{minipage} &
		\begin{minipage}[t]{0.32\hsize}
		\centering
			\includegraphics[width=46mm]{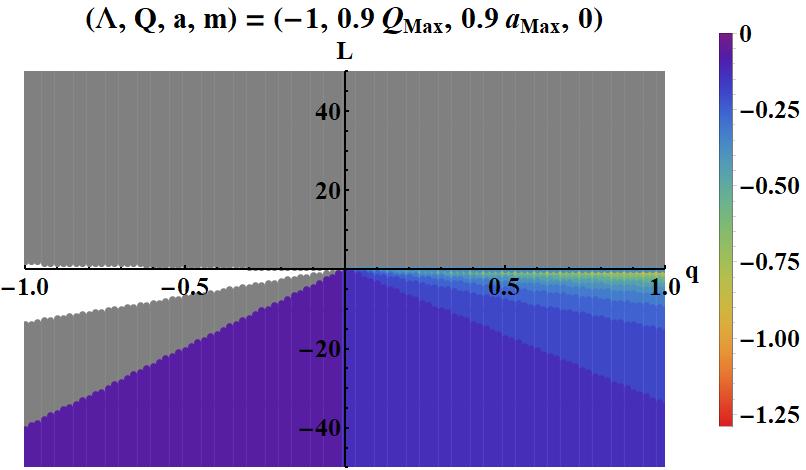}
		\end{minipage} \\
		
		\begin{minipage}[t]{0.32\hsize}
		\centering
			\includegraphics[width=46mm]{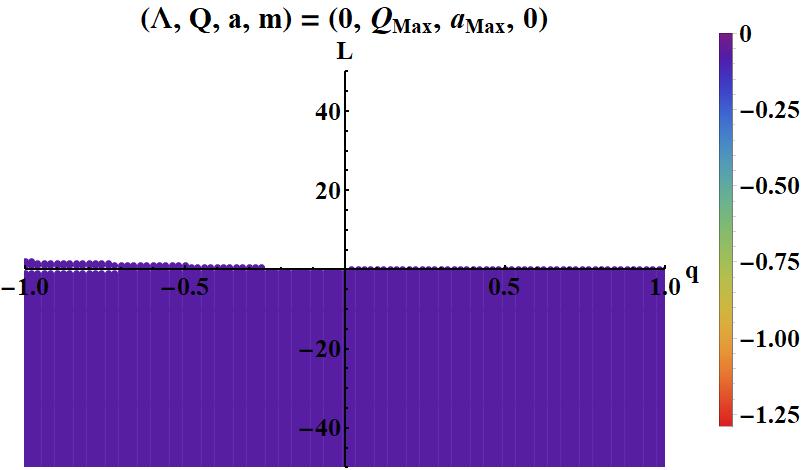}
		\end{minipage} &
		\begin{minipage}[t]{0.32\hsize}
		\centering
			\includegraphics[width=46mm]{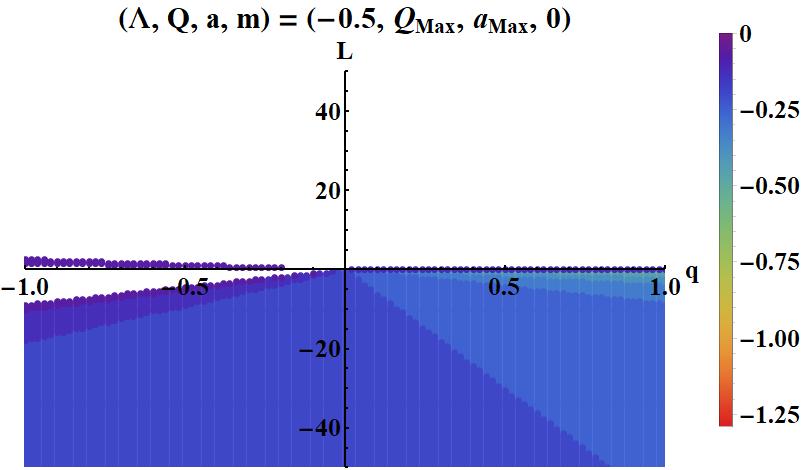}
		\end{minipage} &
		\begin{minipage}[t]{0.32\hsize}
		\centering
			\includegraphics[width=46mm]{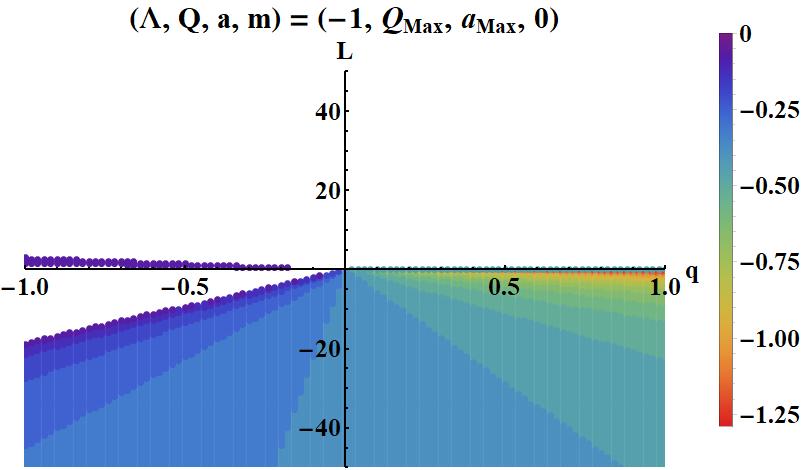}
		\end{minipage} 
	\end{tabular}
	\caption{The result for the particle with $m=0$ in the KN-AdS black hole. \label{fig:m=0_KN}}
\end{figure}
\endgroup

\begingroup
\renewcommand{\arraystretch}{5}
\begin{figure}[H]
	\begin{tabular}{ccc}
		\begin{minipage}[t]{0.32\hsize}
		\centering
			\includegraphics[width=46mm]{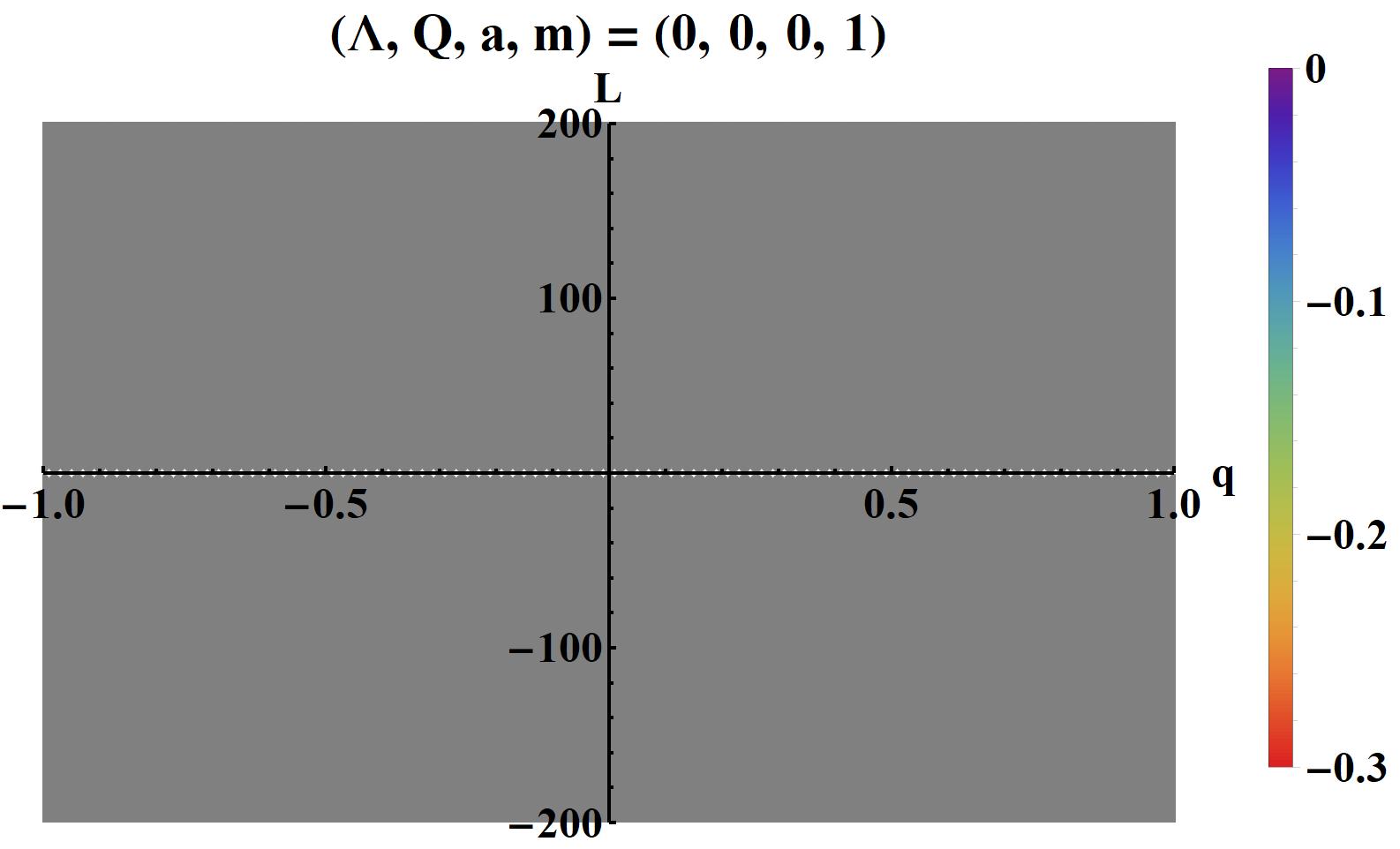}
		\end{minipage} &
		\begin{minipage}[t]{0.32\hsize}
		\centering
			\includegraphics[width=46mm]{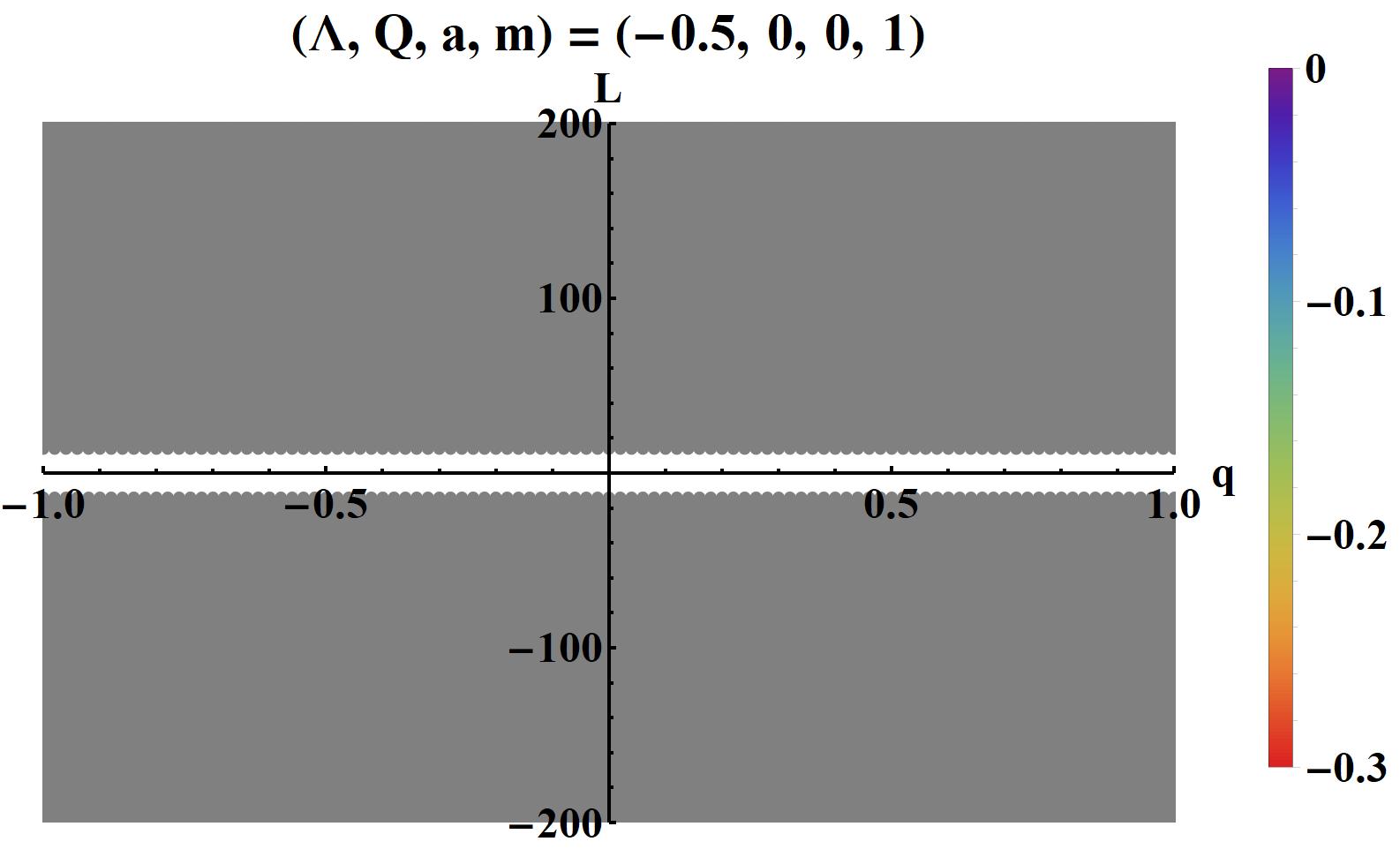}
		\end{minipage} &
		\begin{minipage}[t]{0.32\hsize}
		\centering
			\includegraphics[width=46mm]{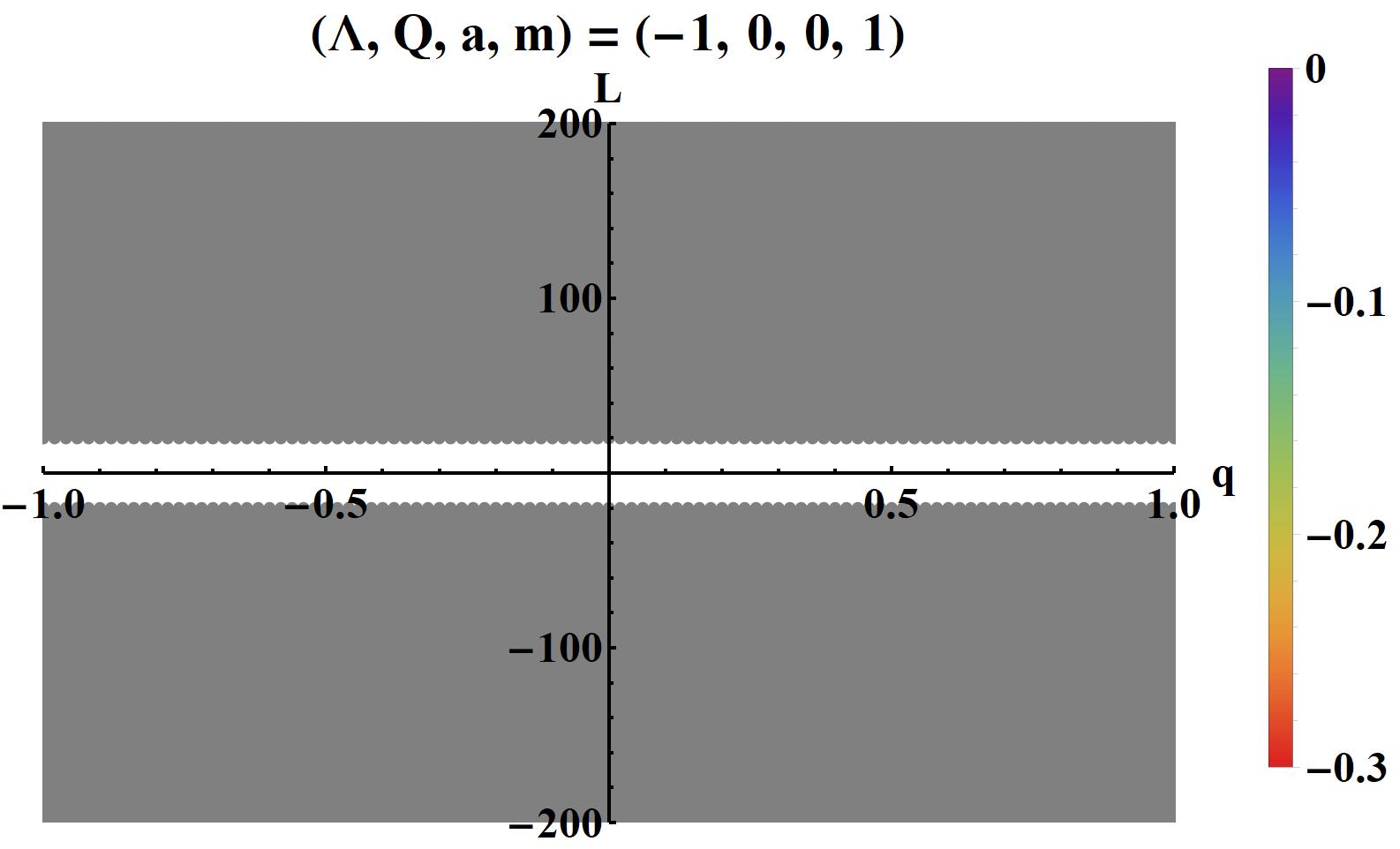}
		\end{minipage} \\

		\begin{minipage}[t]{0.32\hsize}
		\centering
			\includegraphics[width=46mm]{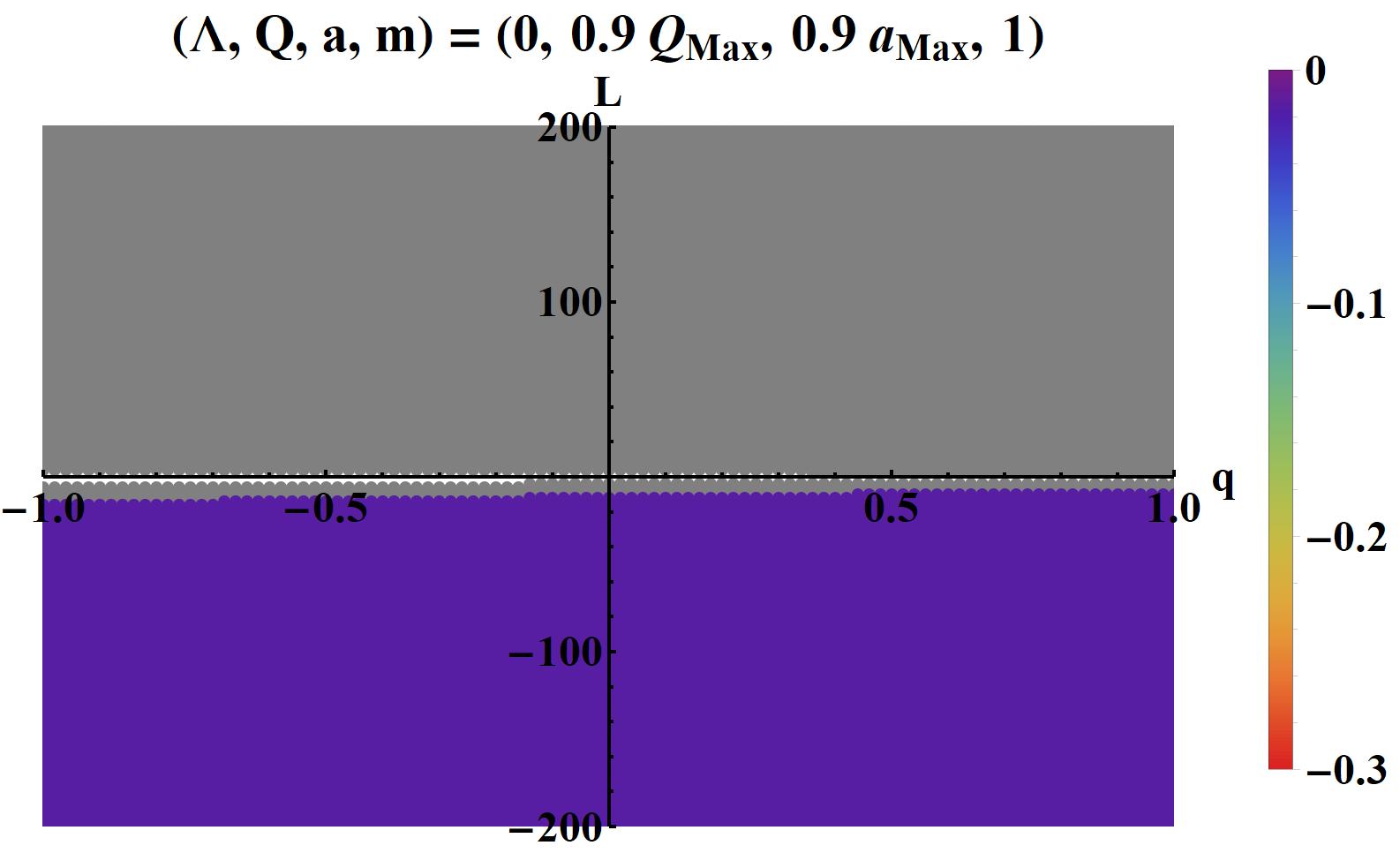}
		\end{minipage} &
		\begin{minipage}[t]{0.32\hsize}
		\centering
			\includegraphics[width=46mm]{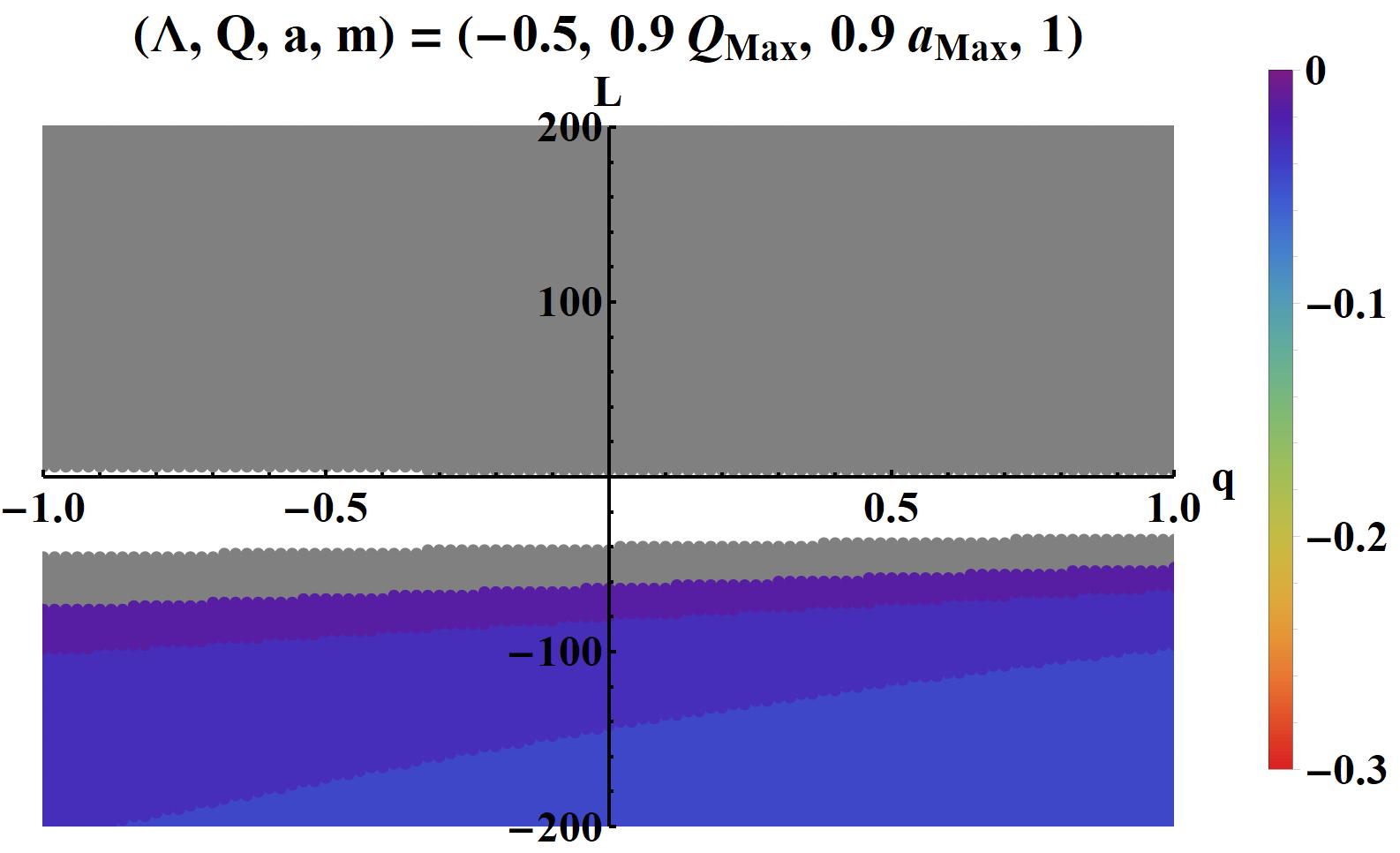}
		\end{minipage} &
		\begin{minipage}[t]{0.32\hsize}
		\centering
			\includegraphics[width=46mm]{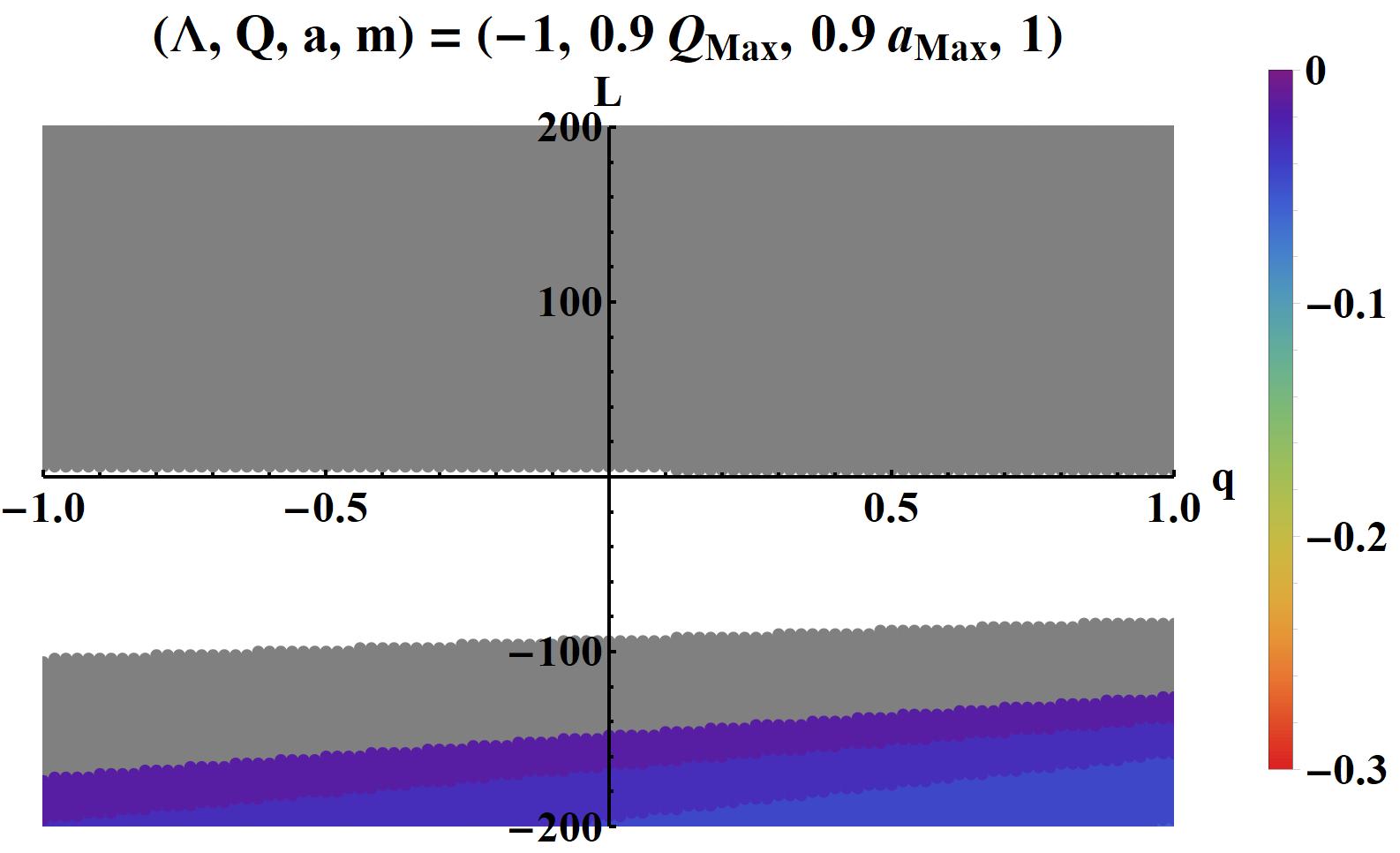}
		\end{minipage} \\
		
		\begin{minipage}[t]{0.32\hsize}
		\centering
			\includegraphics[width=46mm]{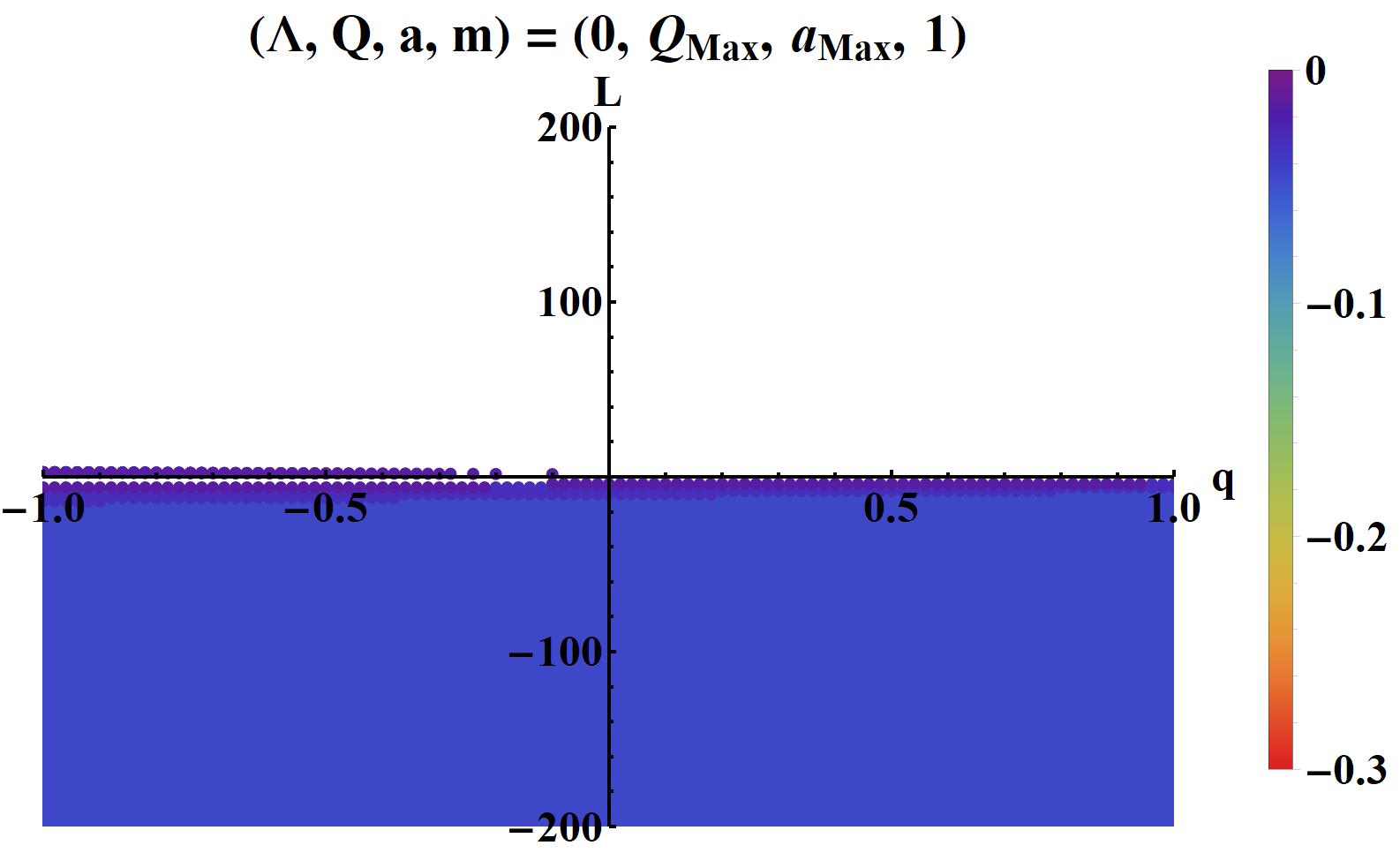}
		\end{minipage} &
		\begin{minipage}[t]{0.32\hsize}
		\centering
			\includegraphics[width=46mm]{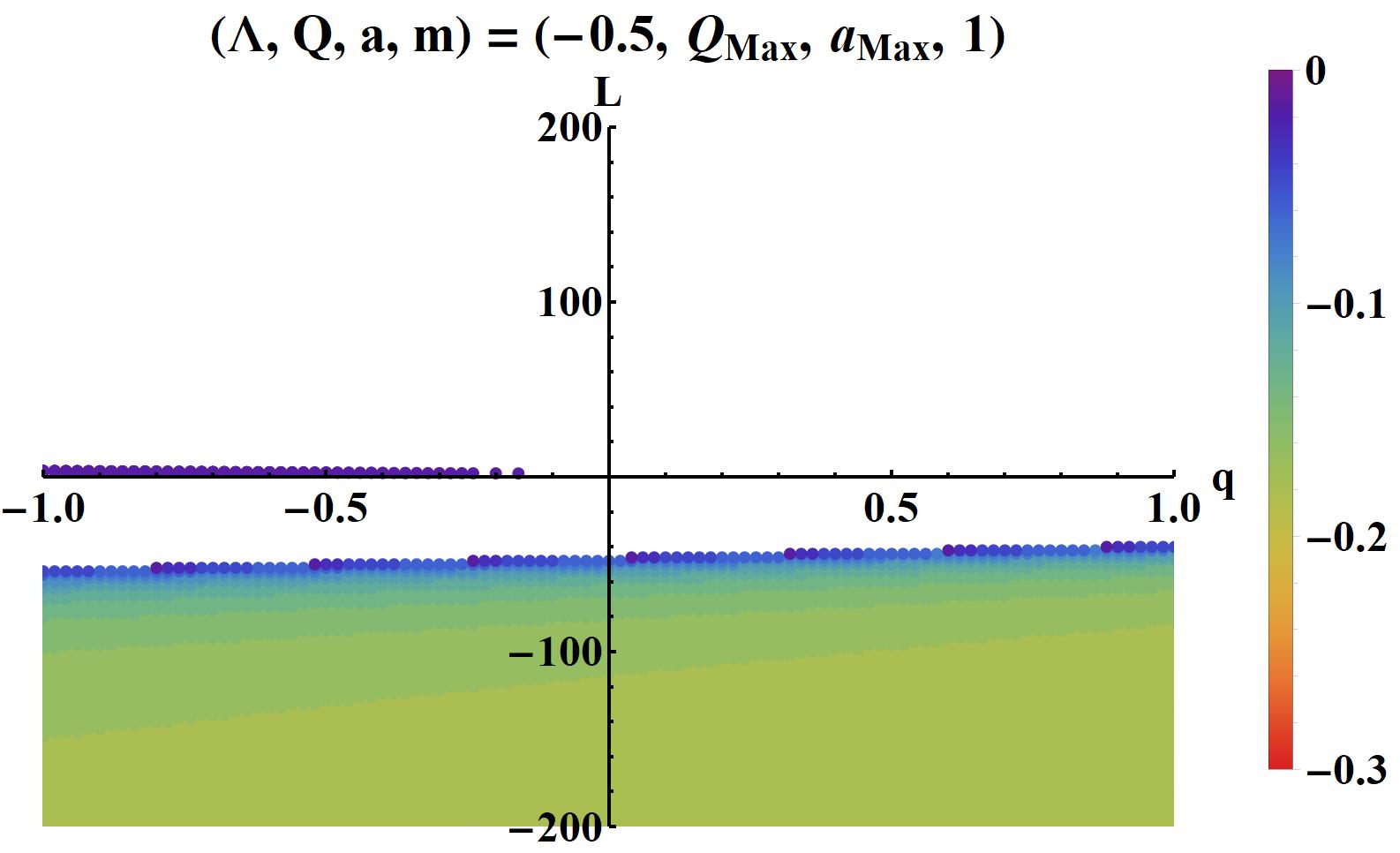}
		\end{minipage} &
		\begin{minipage}[t]{0.32\hsize}
		\centering
			\includegraphics[width=46mm]{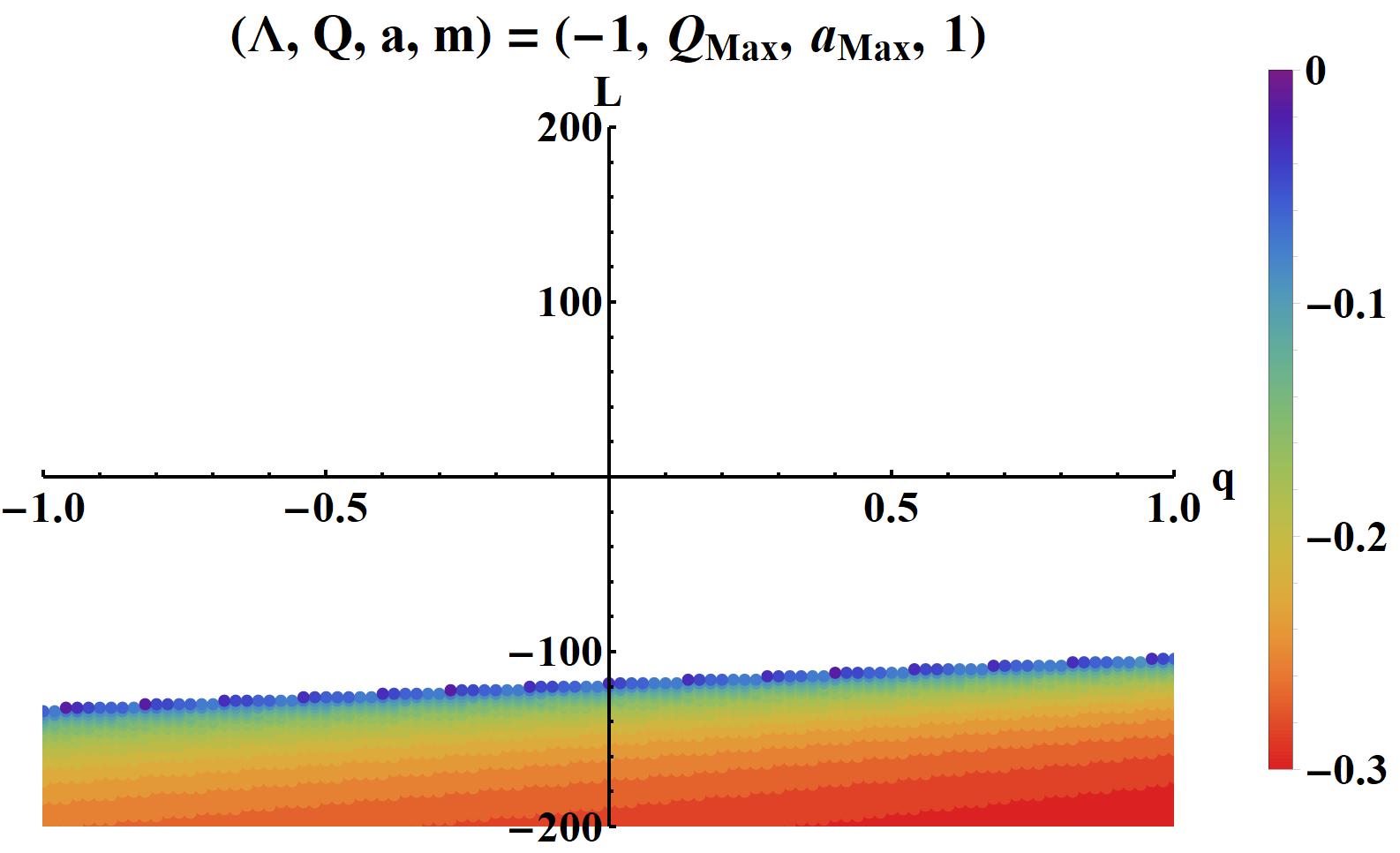}
		\end{minipage} 
	\end{tabular}
	\caption{The result for the particle with $m=1$ in the KN-AdS black hole. \label{fig:m=1_KN}}
\end{figure}
\endgroup

\section{Conclusions}
In this report, we studied the bound on the Lyapunov exponent of a charged probe particle around a black hole with a negative cosmological constant.
We started with the Polyakov-type action of the probe particle with the electric interaction on a curved background.
The Kerr-Newman metric was considered as the background.
Using the static gauge, we found the effective Lagrangian of a particle with angular motion.
The maximum value of the Lyapunov exponent was calculated using the effective Lagrangian.
The bound on the Lyapunov exponent has been conjectured by \cite{Maldacena:2015waa}, which is upper bounded by the surface gravity in the case of a black hole system \cite{Hashimoto:2016dfz}.
We investigated the bound for the AdS black hole including the angular momenta of a particle and a black hole.
It was determined that the bound can be violated when the angular momenta of the particle and/or the black hole have a nonzero value.
This result is remarkable compared to the case where both angular momenta are zero, which was investigated by \cite{Zhao:2018wkl}; for zero angular momenta, the bound is satisfied.

The motion of the particle around the local maximum of the effective potential is described by an inverse harmonic oscillator.
The inverse harmonic potential implies that the motion of the particle in the original Lagrangian is chaotic, and the Lyapunov exponent of the particle is proportional to the second derivative of the effective potential.
The effective potential depends on the parameters of the black hole and the particle, where the contribution of angular momenta, as well as the charges, is important.
Furthermore, compared to the asymptotically-flat case, the value of the Lyapunov exponent drastically changes depending on the cosmological constant.
In this report, we have investigated three types of black holes: the AdS black hole with an electric charge (the RN-AdS black hole), with rotation (the Kerr-AdS black hole), and with an electric charge and rotation (the KN-AdS black hole).
In addition, we investigated massless and massive particles for each type of black hole.
In the case of the RN-AdS black hole, we observed the roles of a negative cosmological constant and the angular momentum of the particle.
It was determined that the negative cosmological constant causes the particle to exhibit less chaotic behavior, whereas the angular momentum of the particle results in chaotic behavior.
In particular, we observed that the bound is violated in most of the parameter regions for the extremal RN-AdS black hole, even if the cosmological constant is negatively large, $\Lambda\sim M$.
In the case of the Kerr-AdS black hole, we analyzed the effect of the angular momentum.
It was determined that when the black hole had a large angular momentum $a$, the particle was more chaotic.
The particle was most chaotic in the extremal limit, whereas in the case of the massless particle, the bound on the Lyapunov exponent was violated.
The results for the KN-AdS black hole exhibited the combined properties of the RN-AdS and Kerr-AdS black holes.

We also investigated the limits of the large negative cosmological constant, near-horizon, and near-extreme in the RN-AdS black hole.
It was demonstrated that $\kappa^2-\lambda^2$ is positive when the cosmological constant is negatively large.
The positivity of $\kappa^2-\lambda^2$ was also shown in the near-horizon limit.
Thus, the bound on the Lyapunov exponent is always satisfied in the limits of a large negative cosmological constant and near-horizon.

In this study, we considered the black holes with the negative cosmological constant.
It is interesting that one consider black holes with the positive cosmological constant, i.e., the dS black holes as a future work.
Besides, it is also interesting that one investigate the bound for black hole solutions in other theories and higher dimensional black holes.


\section*{Acknowledgements}
\noindent 
The work of BG was supported by the National Research Foundation of Korea (NRF) grant funded by the Korea government (MSIT) (NRF-2018R1C1B6004349) and the Dongguk University Research Fund of 2022. 
The work of BHL (NRF-2020R1F1A1075472), and Center for Quantum Spacetime (CQUeST) of Sogang University (NRF-2020R1A6A1A03047877) were supported by Basic Science Research Program through the National Research Foundation of Korea funded by the Ministry of Education.
BG appreciates APCTP for its hospitality during the topical research program, {\it Multi-Messenger Astrophysics and Gravitation}.

\bibliographystyle{jhep}
\bibliography{ref}

\providecommand{\href}[2]{#2}\begingroup\raggedright\begin{thebibliography}{10}

\bibitem{Maldacena:1997re}
J.~M. Maldacena, \emph{{The Large N limit of superconformal field theories and
  supergravity}}, \href{https://doi.org/10.1023/A:1026654312961}{\emph{Adv.
  Theor. Math. Phys.} {\bfseries 2} (1998) 231}
  [\href{https://arxiv.org/abs/hep-th/9711200}{{\ttfamily hep-th/9711200}}].

\bibitem{Gubser:1998bc}
S.~S. Gubser, I.~R. Klebanov and A.~M. Polyakov, \emph{{Gauge theory
  correlators from noncritical string theory}},
  \href{https://doi.org/10.1016/S0370-2693(98)00377-3}{\emph{Phys. Lett. B}
  {\bfseries 428} (1998) 105}
  [\href{https://arxiv.org/abs/hep-th/9802109}{{\ttfamily hep-th/9802109}}].

\bibitem{Witten:1998qj}
E.~Witten, \emph{{Anti-de Sitter space and holography}},
  \href{https://doi.org/10.4310/ATMP.1998.v2.n2.a2}{\emph{Adv. Theor. Math.
  Phys.} {\bfseries 2} (1998) 253}
  [\href{https://arxiv.org/abs/hep-th/9802150}{{\ttfamily hep-th/9802150}}].

\bibitem{Witten:1998zw}
E.~Witten, \emph{{Anti-de Sitter space, thermal phase transition, and
  confinement in gauge theories}},
  \href{https://doi.org/10.4310/ATMP.1998.v2.n3.a3}{\emph{Adv. Theor. Math.
  Phys.} {\bfseries 2} (1998) 505}
  [\href{https://arxiv.org/abs/hep-th/9803131}{{\ttfamily hep-th/9803131}}].

\bibitem{Hawking:1998kw}
S.~W. Hawking, C.~J. Hunter and M.~Taylor, \emph{{Rotation and the AdS / CFT
  correspondence}},
  \href{https://doi.org/10.1103/PhysRevD.59.064005}{\emph{Phys. Rev. D}
  {\bfseries 59} (1999) 064005}
  [\href{https://arxiv.org/abs/hep-th/9811056}{{\ttfamily hep-th/9811056}}].

\bibitem{Larkin:1969qu}
A.~I. Larkin and Y.~N. Ovchinnikov, \emph{{Quasiclassical method in the theory
  of super- conductivity}}, {\emph{JETP} {\bfseries 28} (1969) 1200}.

\bibitem{Shenker:2013pqa}
S.~H. Shenker and D.~Stanford, \emph{{Black holes and the butterfly effect}},
  \href{https://doi.org/10.1007/JHEP03(2014)067}{\emph{JHEP} {\bfseries 03}
  (2014) 067} [\href{https://arxiv.org/abs/1306.0622}{{\ttfamily 1306.0622}}].

\bibitem{Banados:1992wn}
M.~Banados, C.~Teitelboim and J.~Zanelli, \emph{{The Black hole in
  three-dimensional space-time}},
  \href{https://doi.org/10.1103/PhysRevLett.69.1849}{\emph{Phys. Rev. Lett.}
  {\bfseries 69} (1992) 1849}
  [\href{https://arxiv.org/abs/hep-th/9204099}{{\ttfamily hep-th/9204099}}].

\bibitem{Aichelburg:1970dh}
P.~C. Aichelburg and R.~U. Sexl, \emph{{On the Gravitational field of a
  massless particle}}, \href{https://doi.org/10.1007/BF00758149}{\emph{Gen.
  Rel. Grav.} {\bfseries 2} (1971) 303}.

\bibitem{Dray:1984ha}
T.~Dray and G.~'t~Hooft, \emph{{The Gravitational Shock Wave of a Massless
  Particle}}, \href{https://doi.org/10.1016/0550-3213(85)90525-5}{\emph{Nucl.
  Phys. B} {\bfseries 253} (1985) 173}.

\bibitem{Hotta:1992qy}
M.~Hotta and M.~Tanaka, \emph{{Shock wave geometry with nonvanishing
  cosmological constant}},
  \href{https://doi.org/10.1088/0264-9381/10/2/012}{\emph{Class. Quant. Grav.}
  {\bfseries 10} (1993) 307}.

\bibitem{Maldacena:2015waa}
J.~Maldacena, S.~H. Shenker and D.~Stanford, \emph{{A bound on chaos}},
  \href{https://doi.org/10.1007/JHEP08(2016)106}{\emph{JHEP} {\bfseries 08}
  (2016) 106} [\href{https://arxiv.org/abs/1503.01409}{{\ttfamily
  1503.01409}}].

\bibitem{Dettmann:1994dj}
C.~P. Dettmann, N.~E. Frankel and N.~J. Cornish, \emph{{Fractal basins and
  chaotic trajectories in multi - black hole space-times}},
  \href{https://doi.org/10.1103/PhysRevD.50.R618}{\emph{Phys. Rev. D}
  {\bfseries 50} (1994) R618}
  [\href{https://arxiv.org/abs/gr-qc/9402027}{{\ttfamily gr-qc/9402027}}].

\bibitem{Suzuki:1996gm}
S.~Suzuki and K.-i. Maeda, \emph{{Chaos in Schwarzschild space-time: The motion
  of a spinning particle}},
  \href{https://doi.org/10.1103/PhysRevD.55.4848}{\emph{Phys. Rev. D}
  {\bfseries 55} (1997) 4848}
  [\href{https://arxiv.org/abs/gr-qc/9604020}{{\ttfamily gr-qc/9604020}}].

\bibitem{Suzuki:1999si}
S.~Suzuki and K.-i. Maeda, \emph{{Signature of chaos in gravitational waves
  from a spinning particle}},
  \href{https://doi.org/10.1103/PhysRevD.61.024005}{\emph{Phys. Rev. D}
  {\bfseries 61} (2000) 024005}
  [\href{https://arxiv.org/abs/gr-qc/9910064}{{\ttfamily gr-qc/9910064}}].

\bibitem{Han:2008zzf}
W.~Han, \emph{{Chaos and dynamics of spinning particles in Kerr spacetime}},
  \href{https://doi.org/10.1007/s10714-007-0598-9}{\emph{Gen. Rel. Grav.}
  {\bfseries 40} (2008) 1831}
  [\href{https://arxiv.org/abs/1006.2229}{{\ttfamily 1006.2229}}].

\bibitem{Verhaaren:2009md}
C.~Verhaaren and E.~W. Hirschmann, \emph{{Chaotic orbits for spinning particles
  in Schwarzschild spacetime}},
  \href{https://doi.org/10.1103/PhysRevD.81.124034}{\emph{Phys. Rev. D}
  {\bfseries 81} (2010) 124034}
  [\href{https://arxiv.org/abs/0912.0031}{{\ttfamily 0912.0031}}].

\bibitem{PandoZayas:2010xpn}
L.~A. Pando~Zayas and C.~A. Terrero-Escalante, \emph{{Chaos in the Gauge /
  Gravity Correspondence}},
  \href{https://doi.org/10.1007/JHEP09(2010)094}{\emph{JHEP} {\bfseries 09}
  (2010) 094} [\href{https://arxiv.org/abs/1007.0277}{{\ttfamily 1007.0277}}].

\bibitem{Pradhan:2012rkk}
P.~Pradhan, \emph{{Stability analysis and quasinormal modes of
  Reissner\textendash{}Nordstr\o{}m space-time via Lyapunov exponent}},
  \href{https://doi.org/10.1007/s12043-016-1214-x}{\emph{Pramana} {\bfseries
  87} (2016) 5} [\href{https://arxiv.org/abs/1205.5656}{{\ttfamily
  1205.5656}}].

\bibitem{Pradhan:2012qf}
P.~P. Pradhan, \emph{{ISCO, Lyapunov exponent and Kolmogorov-Sinai entropy for
  Kerr-Newman Black hole}},  \href{https://arxiv.org/abs/1212.5758}{{\ttfamily
  1212.5758}}.

\bibitem{Basu:2012ae}
P.~Basu, D.~Das, A.~Ghosh and L.~A. Pando~Zayas, \emph{{Chaos around
  Holographic Regge Trajectories}},
  \href{https://doi.org/10.1007/JHEP05(2012)077}{\emph{JHEP} {\bfseries 05}
  (2012) 077} [\href{https://arxiv.org/abs/1201.5634}{{\ttfamily 1201.5634}}].

\bibitem{Pradhan:2013bli}
P.~P. Pradhan, \emph{{Lyapunov Exponent and Charged Myers Perry Spacetimes}},
  \href{https://doi.org/10.1140/epjc/s10052-013-2477-8}{\emph{Eur. Phys. J. C}
  {\bfseries 73} (2013) 2477}
  [\href{https://arxiv.org/abs/1302.2536}{{\ttfamily 1302.2536}}].

\bibitem{Pradhan:2014tva}
P.~Pradhan, \emph{{Circular Geodesics in Tidal Charged Black Hole}},
  \href{https://doi.org/10.1142/S0219887818500111}{\emph{Int. J. Geom. Meth.
  Mod. Phys.} {\bfseries 15} (2017) 1850011}
  [\href{https://arxiv.org/abs/1412.8123}{{\ttfamily 1412.8123}}].

\bibitem{Jawad:2016kgt}
A.~Jawad, F.~Ali, M.~U. Shahzad and G.~Abbas, \emph{{Dynamics of particles
  around time conformal Schwarzschild black hole}},
  \href{https://doi.org/10.1140/epjc/s10052-016-4422-0}{\emph{Eur. Phys. J. C}
  {\bfseries 76} (2016) 586}
  [\href{https://arxiv.org/abs/1610.05610}{{\ttfamily 1610.05610}}].

\bibitem{Lukes-Gerakopoulos:2016xoc}
G.~Lukes-Gerakopoulos, \emph{{Comment on ''Chaotic orbits for spinning
  particles in Schwarzschild spacetime''}},
  \href{https://doi.org/10.1103/PhysRevD.94.108501}{\emph{Phys. Rev. D}
  {\bfseries 94} (2016) 108501}
  [\href{https://arxiv.org/abs/1604.02955}{{\ttfamily 1604.02955}}].

\bibitem{Chen:2016tmr}
S.~Chen, M.~Wang and J.~Jing, \emph{{Chaotic motion of particles in the
  accelerating and rotating black holes spacetime}},
  \href{https://doi.org/10.1007/JHEP09(2016)082}{\emph{JHEP} {\bfseries 09}
  (2016) 082} [\href{https://arxiv.org/abs/1604.02785}{{\ttfamily
  1604.02785}}].

\bibitem{Hashimoto:2016wme}
K.~Hashimoto, K.~Murata and K.~Yoshida, \emph{{Chaos in chiral condensates in
  gauge theories}},
  \href{https://doi.org/10.1103/PhysRevLett.117.231602}{\emph{Phys. Rev. Lett.}
  {\bfseries 117} (2016) 231602}
  [\href{https://arxiv.org/abs/1605.08124}{{\ttfamily 1605.08124}}].

\bibitem{Jai-akson:2017ldo}
P.~Jai-akson, A.~Chatrabhuti, O.~Evnin and L.~Lehner, \emph{{Black hole merger
  estimates in Einstein-Maxwell and Einstein-Maxwell-dilaton gravity}},
  \href{https://doi.org/10.1103/PhysRevD.96.044031}{\emph{Phys. Rev. D}
  {\bfseries 96} (2017) 044031}
  [\href{https://arxiv.org/abs/1706.06519}{{\ttfamily 1706.06519}}].

\bibitem{Dalui:2018qqv}
S.~Dalui, B.~R. Majhi and P.~Mishra, \emph{{Presence of horizon makes particle
  motion chaotic}},
  \href{https://doi.org/10.1016/j.physletb.2018.11.050}{\emph{Phys. Lett. B}
  {\bfseries 788} (2019) 486}
  [\href{https://arxiv.org/abs/1803.06527}{{\ttfamily 1803.06527}}].

\bibitem{Li:2018wtz}
D.~Li and X.~Wu, \emph{{Chaotic motion of neutral and charged particles in a
  magnetized Ernst-Schwarzschild spacetime}},
  \href{https://doi.org/10.1140/epjp/i2019-12502-9}{\emph{Eur. Phys. J. Plus}
  {\bfseries 134} (2019) 96}
  [\href{https://arxiv.org/abs/1803.02119}{{\ttfamily 1803.02119}}].

\bibitem{Nunez:2018ags}
C.~N\'u\~nez, J.~M. Pen\'\i{}n, D.~Roychowdhury and J.~Van~Gorsel, \emph{{The
  non-Integrability of Strings in Massive Type IIA and their Holographic
  duals}}, \href{https://doi.org/10.1007/JHEP06(2018)078}{\emph{JHEP}
  {\bfseries 06} (2018) 078}
  [\href{https://arxiv.org/abs/1802.04269}{{\ttfamily 1802.04269}}].

\bibitem{Hashimoto:2018fkb}
K.~Hashimoto, K.~Murata and N.~Tanahashi, \emph{{Chaos of Wilson Loop from
  String Motion near Black Hole Horizon}},
  \href{https://doi.org/10.1103/PhysRevD.98.086007}{\emph{Phys. Rev. D}
  {\bfseries 98} (2018) 086007}
  [\href{https://arxiv.org/abs/1803.06756}{{\ttfamily 1803.06756}}].

\bibitem{Akutagawa:2018yoe}
T.~Akutagawa, K.~Hashimoto, T.~Miyazaki and T.~Ota, \emph{{Phase diagram of QCD
  chaos in linear sigma models and holography}},
  \href{https://doi.org/10.1093/ptep/pty055}{\emph{PTEP} {\bfseries 2018}
  (2018) 063B01} [\href{https://arxiv.org/abs/1804.01737}{{\ttfamily
  1804.01737}}].

\bibitem{Akutagawa:2019awh}
T.~Akutagawa, K.~Hashimoto, K.~Murata and T.~Ota, \emph{{Chaos of QCD string
  from holography}},
  \href{https://doi.org/10.1103/PhysRevD.100.046009}{\emph{Phys. Rev. D}
  {\bfseries 100} (2019) 046009}
  [\href{https://arxiv.org/abs/1903.04718}{{\ttfamily 1903.04718}}].

\bibitem{Cubrovic:2019qee}
M.~\v{C}ubrovi\'c, \emph{{The bound on chaos for closed strings in Anti-de
  Sitter black hole backgrounds}},
  \href{https://doi.org/10.1007/JHEP12(2019)150}{\emph{JHEP} {\bfseries 12}
  (2019) 150} [\href{https://arxiv.org/abs/1904.06295}{{\ttfamily
  1904.06295}}].

\bibitem{Giataganas:2021ghs}
D.~Giataganas, \emph{{Chaotic Motion near Black Hole and Cosmological
  Horizons}},  \href{https://arxiv.org/abs/2112.02081}{{\ttfamily 2112.02081}}.

\bibitem{Lei:2021koj}
Y.-Q. Lei and X.-H. Ge, \emph{{Circular Motion of Charged Particles near
  Charged Black Hole}},  \href{https://arxiv.org/abs/2111.06089}{{\ttfamily
  2111.06089}}.

\bibitem{Barrow:1981sx}
J.~D. Barrow, \emph{{Chaotic behavior in general relativity}},
  \href{https://doi.org/10.1016/0370-1573(82)90171-5}{\emph{Phys. Rept.}
  {\bfseries 85} (1982) 1}.

\bibitem{Bombelli:1991eg}
L.~Bombelli and E.~Calzetta, \emph{{Chaos around a black hole}},
  \href{https://doi.org/10.1088/0264-9381/9/12/004}{\emph{Class. Quant. Grav.}
  {\bfseries 9} (1992) 2573}.

\bibitem{Yurtsever:1994yb}
U.~Yurtsever, \emph{{Geometry of chaos in the two center problem in general
  relativity}}, \href{https://doi.org/10.1103/PhysRevD.52.3176}{\emph{Phys.
  Rev. D} {\bfseries 52} (1995) 3176}
  [\href{https://arxiv.org/abs/gr-qc/9412031}{{\ttfamily gr-qc/9412031}}].

\bibitem{Letelier:1997uv}
P.~S. Letelier and W.~M. Vieira, \emph{{Chaos and rotating black holes with
  halos}}, \href{https://doi.org/10.1103/PhysRevD.56.8095}{\emph{Phys. Rev. D}
  {\bfseries 56} (1997) 8095}
  [\href{https://arxiv.org/abs/gr-qc/9712008}{{\ttfamily gr-qc/9712008}}].

\bibitem{deMoura:1999wf}
A.~P.~S. de~Moura and P.~S. Letelier, \emph{{Chaos and fractals in geodesic
  motions around a nonrotating black hole with an external halo}},
  \href{https://doi.org/10.1103/PhysRevE.61.6506}{\emph{Phys. Rev. E}
  {\bfseries 61} (2000) 6506}
  [\href{https://arxiv.org/abs/chao-dyn/9910035}{{\ttfamily
  chao-dyn/9910035}}].

\bibitem{Setare:2010zd}
M.~R. Setare and D.~Momeni, \emph{{Geodesic stability for KS Black hole in
  Horava-Lifshitz gravity via Lyapunov exponents}},
  \href{https://doi.org/10.1007/s10773-010-0498-8}{\emph{Int. J. Theor. Phys.}
  {\bfseries 50} (2011) 106} [\href{https://arxiv.org/abs/1001.3767}{{\ttfamily
  1001.3767}}].

\bibitem{Lukes-Gerakopoulos:2016udm}
G.~Lukes-Gerakopoulos, \emph{{Spinning particles moving around black holes:
  integrability and chaos}},  in \emph{{14th Marcel Grossmann Meeting on Recent
  Developments in Theoretical and Experimental General Relativity,
  Astrophysics, and Relativistic Field Theories}}, vol.~2, pp.~1960--1965,
  2017, \href{https://arxiv.org/abs/1606.09430}{{\ttfamily 1606.09430}},
  \href{https://doi.org/10.1142/9789813226609_0209}{DOI}.

\bibitem{Liu:2017fjx}
C.-Y. Liu, D.-S. Lee and C.-Y. Lin, \emph{{Geodesic motion of neutral particles
  around a Kerr\textendash{}Newman black hole}},
  \href{https://doi.org/10.1088/1361-6382/aa903b}{\emph{Class. Quant. Grav.}
  {\bfseries 34} (2017) 235008}
  [\href{https://arxiv.org/abs/1706.05466}{{\ttfamily 1706.05466}}].

\bibitem{Zelenka:2019nyp}
O.~Zelenka, G.~Lukes-Gerakopoulos, V.~Witzany and O.~Kop\'a\v{c}ek,
  \emph{{Growth of resonances and chaos for a spinning test particle in the
  Schwarzschild background}},
  \href{https://doi.org/10.1103/PhysRevD.101.024037}{\emph{Phys. Rev. D}
  {\bfseries 101} (2020) 024037}
  [\href{https://arxiv.org/abs/1911.00414}{{\ttfamily 1911.00414}}].

\bibitem{Yi:2020shw}
M.~Yi and X.~Wu, \emph{{Dynamics of charged particles around a magnetically
  deformed Schwarzschild black hole}},
  \href{https://doi.org/10.1088/1402-4896/aba4c2}{\emph{Phys. Scripta}
  {\bfseries 95} (2020) 085008}.

\bibitem{Mondal:2021exj}
M.~Mondal, F.~Rahaman and K.~N. Singh, \emph{{Lyapunov exponent ISCO and
  Kolmogorov Senai entropy for Kerr Kiselev black hole}},
  \href{https://doi.org/10.1140/epjc/s10052-021-08888-1}{\emph{Eur. Phys. J. C}
  {\bfseries 81} (2021) 84} [\href{https://arxiv.org/abs/2102.02667}{{\ttfamily
  2102.02667}}].

\bibitem{Hashimoto:2016dfz}
K.~Hashimoto and N.~Tanahashi, \emph{{Universality in Chaos of Particle Motion
  near Black Hole Horizon}},
  \href{https://doi.org/10.1103/PhysRevD.95.024007}{\emph{Phys. Rev. D}
  {\bfseries 95} (2017) 024007}
  [\href{https://arxiv.org/abs/1610.06070}{{\ttfamily 1610.06070}}].

\bibitem{Zhao:2018wkl}
Q.-Q. Zhao, Y.-Z. Li and H.~Lu, \emph{{Static Equilibria of Charged Particles
  Around Charged Black Holes: Chaos Bound and Its Violations}},
  \href{https://doi.org/10.1103/PhysRevD.98.124001}{\emph{Phys. Rev. D}
  {\bfseries 98} (2018) 124001}
  [\href{https://arxiv.org/abs/1809.04616}{{\ttfamily 1809.04616}}].

\bibitem{Lei:2020clg}
Y.-Q. Lei, X.-H. Ge and C.~Ran, \emph{{Chaos of particle motion near a black
  hole with quasitopological electromagnetism}},
  \href{https://doi.org/10.1103/PhysRevD.104.046020}{\emph{Phys. Rev. D}
  {\bfseries 104} (2021) 046020}
  [\href{https://arxiv.org/abs/2008.01384}{{\ttfamily 2008.01384}}].

\bibitem{Kan:2021blg}
N.~Kan and B.~Gwak, \emph{{Bound on the Lyapunov exponent in Kerr-Newman black
  holes via a charged particle}},
  \href{https://doi.org/10.1103/PhysRevD.105.026006}{\emph{Phys. Rev. D}
  {\bfseries 105} (2022) 026006}
  [\href{https://arxiv.org/abs/2109.07341}{{\ttfamily 2109.07341}}].

\bibitem{Yu:2022ysm}
C.~Yu, D.~Chen and C.~Gao, \emph{{The bound of Lyapunov exponent in
  Einstein-Maxwell-Dilaton-Axion black holes}},
  \href{https://arxiv.org/abs/2202.13741}{{\ttfamily 2202.13741}}.

\bibitem{Caldarelli:1998hg}
M.~M. Caldarelli and D.~Klemm, \emph{{Supersymmetry of Anti-de Sitter black
  holes}}, \href{https://doi.org/10.1016/S0550-3213(98)00846-3}{\emph{Nucl.
  Phys. B} {\bfseries 545} (1999) 434}
  [\href{https://arxiv.org/abs/hep-th/9808097}{{\ttfamily hep-th/9808097}}].

\bibitem{Morita:2021syq}
T.~Morita, \emph{{Extracting Classical Lyapunov Exponent from One-Dimensional
  Quantum Mechanics}},  \href{https://arxiv.org/abs/2105.09603}{{\ttfamily
  2105.09603}}.

\bibitem{Morita:2021mfi}
T.~Morita, \emph{{Analogous Hawking Radiation in Butterfly Effect}},  in
  \emph{{4th International Conference on Holography, String Theory and Discrete
  Approach in Hanoi, Vietnam}}, 1, 2021,
  \href{https://arxiv.org/abs/2101.02435}{{\ttfamily 2101.02435}}.

\bibitem{Bhattacharyya:2020art}
A.~Bhattacharyya, W.~Chemissany, S.~S. Haque, J.~Murugan and B.~Yan, \emph{{The
  Multi-faceted Inverted Harmonic Oscillator: Chaos and Complexity}},
  \href{https://doi.org/10.21468/SciPostPhysCore.4.1.002}{\emph{SciPost Phys.
  Core} {\bfseries 4} (2021) 002}
  [\href{https://arxiv.org/abs/2007.01232}{{\ttfamily 2007.01232}}].

\end{thebibliography}\endgroup
\end{document}